\documentclass[12pt,notitlepage]{article}
\pdfoutput=1

\usepackage{fouriernc}
\usepackage[utf8]{inputenc}

\usepackage{amsfonts}
\usepackage{amssymb}
\usepackage{amstext} 
\usepackage{amsmath}

\usepackage{xspace}
\usepackage{theorem}
\usepackage{url}
\usepackage{graphicx}

\usepackage{float} 
\usepackage{listings}%
\usepackage{xcolor}
\definecolor{bg}{RGB}{255,255,226}
\definecolor{mylilas}{RGB}{170,55,241} 

\usepackage[letterpaper,bindingoffset=0.2in,%
            left=1in,right=1in,top=1in,bottom=1in,%
            footskip=.25in]{geometry}
\usepackage{hyperref}
\begin{document}
\begin{titlepage}
\thispagestyle{empty}

\title{Analysis and object oriented implementation\\of the Kovacic algorithm}
\author{Nasser M. Abbasi\thanks{\href{https://12000.org/}{https://12000.org/}}}
\date{}
\maketitle

\vfill

\begin{abstract}
This paper gives a detailed overview and a number of worked out examples illustrating the 
Kovacic~\cite{Kovacic86} algorithm for solving second order linear differential 
equation ${A(x) y''+ B(x) y' + C(x) y=0}$ where $A,B,C$ are rational functions with 
complex coefficients in the independent variable $x$. All three cases of the algorithm were
implemented in a software package based on an object oriented design and complete source 
code listing given in the appendix with usage examples. Implementation used the Maple 
computer algebra language.\footnote{The complete Kovacic package in one mpl file accompany the arXiv version of this paper}.
This package was then used to analyze the distribution of Kovacic algorithm cases on $3000$ 
differential equations.
\end{abstract}

\vfill 

\renewcommand{\section}[2]{}

\setcounter{tocdepth}{2} 
\tableofcontents

\vfill

\end{titlepage}

\section{Introduction}
Kovacic~\cite{Kovacic86} gave an algorithm for finding a closed form Liouvillian\footnote{Wikipedia defines
Liouvillian function as function of one variable which is the composition of a finite number of 
arithmetic operations $(+,-, \times, \div )$, exponentials, constants, solutions of algebraic equations (a 
generalization of nth roots), and antiderivatives. Kovacic in his original paper says 
``Such a solution may involve exponentials, indefinite integrals and
solutions of polynomial equations. (As we are considering functions of a complex
variable, we need not explicitly mention trigonometric functions, they can be written in
terms of exponentials. Note that logarithms are indefinite integrals and hence are
allowed.''} solution to any linear second order differential equation ${A y''+ B y' + C y=0}$ 
if such a solution exists. Smith~\cite{Smith84}  gave an implementation 
based on a modified version of Kovacic algorithm by Saunders~\cite{Saunders}. 

The current implementation is based on the original paper by Kovacic and uses the new 
object oriented features in Maple. The accompanied software package have been 
tested on $3000$ differential equations with each solution verified using Maple's odetest. The test 
suite is included as a separate module. The Appendix describes how to use the software.

The Kovacic algorithm finds one (basis) solution of ${A y''+ B y' + C y=0}$. The second basis solution
is found using reduction of order. The general solution is a linear combination of
the two basis solutions found.

The algorithm starts by writing  the input ode $A y''+ B y' + C y=0$ as 
\begin{align*}
y''+ a y' + b y=0  \tag{1}
\end{align*}
Where $a=\frac{B}{A},b=\frac{C}{A}$. The substitution 
\begin{align*}
z &= y e^{\frac{1}{2}\int a\,dx} \tag{2} 
\end{align*}
is then applied to (1) which transforms it to a second order ode in the new dependent 
variable $z(x)$ without the first derivative
\begin{align*} 
     z''&=r z \tag{3} 
\end{align*} 
$r$ in the above is given by 
\begin{align*} 
 r &= \frac{1}{4} a^2 + \frac{1}{2} a' - b \tag{4} 
\end{align*} 
It is ode (3) which is solved by the algorithm and not (1). Equation (3) will be called the DE from now on. 

If a solution $z(x)$ to the DE is found, then the first basis solution to the original ode
is obtained using the transformation (2) in reverse
\begin{align*}
y &= z e^{-\frac{1}{2}\int a\,dx}
\end{align*}
The second solution is found using reduction of order.

\begin{center}
\begin{minipage}{\textwidth}
These are the four possible cases to consider.
\begin{enumerate}
\item DE has solution $z=e^{\int\omega dx}$ with $\omega\in \mathbb{C}(x)$.

\item DE has solution $z=e^{\int\omega dx}$ with $\omega$  polynomial over
$\mathbb{C}(x)$ of degree $2$.
 
\item Solutions of DE are algebraic over $\mathbb{C}(x)$.

\item DE has no Liouvillian solution.
\end{enumerate}
\end{minipage}
\end{center}

Before describing how the algorithm works, there are necessary (but not
sufficient) conditions that determine which case the DE satisfies.
Only those cases that meet the necessary conditions will be attempted.

The following are the necessary conditions for each case. To check each
case, let $r=\frac{s}{t}$ where $\gcd(s,t) =1$. The order of $r$ at $\infty$ (from now on
referred to as  $\mathcal{O}(\infty)$) is defined as $\deg(t)-\deg(s)$. The poles of $r$ and the 
order of each pole need to be determined.

Knowing the order of the poles of $r$ and $\mathcal{O}(\infty)$ is all what 
is needed to determine the necessary conditions for each case. These conditions are
the following

\begin{center}
\begin{minipage}{\textwidth}
\begin{enumerate}
\item Case $1$. Either no pole exists, or if a pole exists, the order must be either one or even. 
If $\mathcal{O}(\infty)$ is less $3$, then it must be even otherwise it can be even or odd.

\item Case $2$. $r$ must have at least one pole either of order 2 or odd order greater than 2. 
There are no conditions on $\mathcal{O}(\infty)$.

\item Case $3$. $r$ must have a pole either of order $1$ or $2$. No other order is allowed. 
$\mathcal{O}(\infty)$ must be at least $2$.
\end{enumerate}
\end{minipage}
\end{center}

If the conditions of a case are not satisfied then the case will be attempted
as the algorithm guarantees that there will be no Liouvillian solution. However if the conditions 
are satisfied, this does not necessarily mean a solution exists. As an 
example $y''=1/x^6 y$ satisfies only case one, but running the algorithm on case one
shows that there is no Liouvillian solution. 

The following table summarizes the above conditions for each case.
 
\begin{table}[H]
\centering
\begin{tabular}[c]{|p{.3in}|p{2.5in}|p{2.5in}|}\hline
Case & Allowed pole order for $r$ & Allowed value for $\mathcal{O}(\infty)$ \\\hline
1 & $\left\{  0,1,2,4,6,8,\cdots\right\}  $ & $\left\{  \cdots,-6,-4,-2,0,2,3,4,5,6,\cdots\right\}  $ \\\hline
2 & Need to have at least one pole of order $2$ or pole of odd order greater than $2$. Any other pole
order is allowed as long as the above condition is satisfied. The following are examples of pole 
orders which are allowed. $\{1,2\}$,$\{1,3\}$,$\{2\}$,$\{3\}$,$\{3,4\}$,$\{1,2,5\}$.  & no conditions \\\hline
3 & $\left\{  1,2\right\}  $ & $\left\{  2,3,4,5,6,7,\cdots\right\}  $ \\\hline
\end{tabular}
\caption{Necessary conditions for each Kovacic case}\label{tab:first}
\end{table}

Some observations: In case one, no odd order pole is allowed except for order 1. 
Case one is the only case that could have no pole in $r$, which is the same as a pole of order zero. 
Case two and three require at least one pole. For case three, only poles of order $1$ or $2$ are allowed. 
If  $\mathcal{O}(\infty)$ is zero, then only possibility is either case one or two. 
For case one, if $\mathcal{O}(\infty)$ is negative, then it must be even. 

The above table also shows that when $r$ has only one pole of order $2$ and $\mathcal{O}(\infty)$ 
equals $2$ or higher then all three cases are possible. Also, if $r$ has 
two poles one of order $1$ and the other of order $2$ and $\mathcal{O}(\infty)$ equals $2$ or higher 
then all three cases are possible. 

These are the only two possibilities where all three cases have the same necessary conditions.

\section{Description of algorithm for each case}
\subsection{Case one}
\subsubsection{step 1}  
Assuming that the necessary conditions for case one are satisfied 
and $z''=r z, r=\frac{s}{t}$. Let $\Gamma$  be the set of all poles 
of $r$. For each pole $c$ in this set, three quantities are calculated:~Rational
function $\left[  \sqrt{r}\right]_{c}$ and two complex numbers $\alpha_{c}^{+},\alpha_{c}^{-}$. 

How this is done depends on the order of the pole as described below. 
If the set $\Gamma$  is  empty (when there are no poles), then this part is skipped.
\begin{enumerate}
\item If the pole $c$ has order $1$ then
\begin{align*}
\left[  \sqrt{r}\right]  _{c}  &  =0\\
\alpha_{c}^{+}  &  =1\\
\alpha_{c}^{-}  &  =1
\end{align*}
\item If the pole $c$ is of order $2$ then
\begin{align*}
\left[  \sqrt{r}\right]_{c}  &  =0\\
\alpha_{c}^{+}  &  =\frac{1}{2}+\frac{1}{2}\sqrt{1+4b}\\
\alpha_{c}^{-}  &  =\frac{1}{2}-\frac{1}{2}\sqrt{1+4b}
\end{align*}
Where $b$ is the coefficient of $\frac{1}{(x-c)^{2}}$ in the partial fraction decomposition of $r$.
\item If the pole is of order $\{4,6,8,\dots\}$ (poles must be all even
from the conditions of case one), then the computation is 
more involved. Let $2v$ be the order of the pole. Hence if the pole was 
order 4, then $v=2$. Let $\left[  \sqrt{r}\right]_{c}$ be the sum of 
terms involving $\frac{1}{(x-c)^i}$ for $2\leq i\leq v$ in the Laurent series
expansion of $\sqrt{r}$ (not $r$) at $c$. Therefore
\begin{align*}
\left[  \sqrt{r}\right]  _{c}  &=\sum_{i=2}^{v}\frac{a_{i}}{\left(x-c\right)  ^{i}} \\
       &= \frac{a_{2}}{\left(  x-c\right)  ^{2}}+\frac{a_{3}}{\left(  x-c\right) ^{3}}+\cdots+\frac{a_{v}}{\left(  x-c\right)  ^{v}}\tag{1}
\end{align*}
$\alpha_{c}^{+},\alpha_{c}^{-}$ are found using
\begin{align*} 
\alpha_{c}^{+}  &  =\frac{1}{2}\left(  \frac{b}{a_v}+v\right)\\
\alpha_{c}^{-}  &  =\frac{1}{2}\left(  -\frac{b}{a_v}+v\right)
\end{align*}
Where in the above $a_v$ is the coefficient of the term $\frac{a_v}{(x-c)^v}$ in (1)
and $b$ is the coefficient of the term $\frac{1}{(x-c)^{v+1}}$ in $r$ itself (found
from the partial fraction decomposition),  minus the coefficient 
of same term in the Laurent series expansion of $\sqrt{r}$ at $c$.

The coefficients in the Laurent series can be obtained as follows. Given $r(x)$ with
a pole of finite order $N$ at $x=c$, then its Laurent series expansion at $c$ is given by
the sum of the analytic part and the principal part of the of the Laurent series. The
coefficients $b_n$ are contained in the principal part of the series.
\begin{align}
r\left(  x\right)   &  =\sum_{n=0}^{\infty}a_{n}\left(  x-c\right)^{n}+
\sum_{n=1}^{N}\frac{b_{n}}{\left(  x-c\right)  ^{n}}\tag{2}\\
&  =\sum_{n=0}^{\infty}a_{n}\left(  x-c\right)  ^{n}+\frac{b_{1}}{\left(
x-c\right)  }+\frac{b_{2}}{\left(  x-c\right)  ^{2}}+\frac{b_{3}}{\left(  x-c\right)  ^{3}}
+\cdots+\frac{b_{N}}{\left(  x-c\right)^{N}}\nonumber
\end{align}
To obtain $b_{1}$ (which is the residue of $r\left(  x\right)  $ at $c$), 
both sides of the above are multiplied by $\left(x-c\right)^{N}$ which gives
\begin{equation}
\left(  x-c\right)  ^{N}r\left(  x\right)  =\sum_{n=0}^{\infty}%
a_{n}\left(  x-c\right)  ^{n+N}+b_{1}\left(  x-c\right)  ^{N-1}%
+b_{2}\left(  x-c\right)  ^{N-2}+\cdots+b_{N}\tag{3}%
\end{equation}
Differentiating both sides of (3) $(N-1)$ times w.r.t. $x$ gives%
\[
\frac{d^{N-1}}{dx^{\left(  N-1\right)  }}\left(  \left(  x-c\right)^{N}f\left(  x\right)  \right)  
  =\sum_{n=0}^{\infty}\frac{d^{N-1}}{dx^{\left(N-1\right)  }}\left(  a_{n}\left(  x-c\right)  ^{n+N}\right)
+b_{1}\left(  N-1\right)  !
\]
Evaluating the above at $x=c$ gives%
\[
b_{1}=\frac{\lim_{x\rightarrow c}\frac{d^{N-1}}{dx^{\left(  N-1\right)  }%
}\left(  \left(  x-c\right)  ^{N}r(x)  \right)  }{\left(
N-1\right)  !}%
\]
To find the next coefficient $b_{2}$, both sides of (3) are differentiated $(N-2)$ times
\[
\frac{d^{N-2}}{dx^{\left(  N-2\right)  }}\left(  \left(  x-c\right)
^{N}r\left(  x\right)  \right)  =\sum_{n=0}^{\infty}\frac{d^{N-2}}{dx^{\left(
N-2\right)  }}\left(  a_{n}\left(  x-c\right)  ^{n+N}\right)
+b_{1}\left(  N-1\right)  !\left(  x-c\right)  +b_{2}\left(  N-2\right)  !
\]
Evaluating the above at $x=c$ gives%
\[
b_{2}=\frac{\lim_{x\rightarrow c}\frac{d^{N-2}}{dx^{\left(  N-2\right)  }%
}\left(  \left(  x-c\right)  ^{N}r\left(  x\right)  \right)  }{\left(
N-2\right)  !}%
\]
The above is repeated to find $b_{3},b_{4},\cdots,b_{N}$. The
general formula for find coefficient $b_n$ is therefore
\begin{align*}
b_{n}=\frac{\lim_{x\rightarrow c}\frac{d^{N-n}}{dx^{\left(  N-n\right)  }%
}\left(  \left(  x-c\right)  ^{N}r\left(  x\right)  \right)  }{\left(
N-n\right)  !}\tag{4} 
\end{align*}
For the special case of the last term $b_{N}$ the above simplifies to%
\begin{align*}
b_{N}=\lim_{x\rightarrow c}(x-c)^{N} r(x) \tag{5} 
\end{align*}
The above is implemented in the function \verb|laurent_coeff()| in the Kovacic class.

This completes finding all the quantities 
$\left\{\left[  \sqrt{r}\right]  _{c}, \alpha_{c}^{+},\alpha_{c}^{+}\right\}$ for each pole in the set
$\Gamma$ for case one.
\end{enumerate}

The next step calculates the following three quantities for $\mathcal{O}(\infty)$. 
\begin{enumerate}
\item If $\mathcal{O}(\infty)\leq 0$, which must be even, then let $-2 v =\mathcal{O}(\infty)$ and
 $\left[  \sqrt{r}\right]  _{\infty}$ is then the sum of all terms $x^i$ for for $0\leq i\leq v$ in
the Laurent series expansion of $\sqrt{r}$ at $\infty$.
\begin{align*} 
\left[  \sqrt{r}\right]  _{\infty}&= \sum_{i=0}^{v} a_i x^v  = a_0 + a_1 x + a_2 x^2 \cdots+ a_{v} x^v \tag{6} 
\end{align*} 
The coefficients $a_i$ are found by setting $x=\frac{1}{y}$ in $r$ and then finding the Laurent series
of $\left[  \sqrt{r(y)}\right]$  expanded around $y=zero$. The process for finding the 
coefficient is the same one used as described earlier where now the limit is taken 
as $y$ approaches zero from the right. This gives all the terms of (6). This 
is implemented in the function \verb|laurent_coeff()| in the Kovacic class.
 
The corresponding $\{\alpha_{\infty}^{+},\alpha_{\infty}^{-}\}$ are given by
\begin{align*}
\alpha_{\infty}^{+}  &  =\frac{1}{2}\left(  \frac{b}{a_v}-v\right) \\
\alpha_{\infty}^{-}  &  =\frac{1}{2}\left(  -\frac{b}{a_v}-v\right)
\end{align*}
Where $a_v$  is coefficient of $x^v$ in (6) and $b$ is the coefficient of $x^{v-1}$ in $r$ itself
(found using long division) minus the coefficient of  
$x^{v-1}$ in $\left(\left[  \sqrt{r}\right]  _{\infty}\right)^2$.
\item If $\mathcal{O}(\infty)=2$ then $\left[  \sqrt{r}\right]  _{\infty}=0$. The 
corresponding $\{\alpha_{\infty}^{+},\alpha_{\infty}^{-}\}$ are given by
\begin{align*}
\alpha_{\infty}^{+}  &  =\frac{1}{2}+\frac{1}{2}\sqrt{1+4b}\\
\alpha_{\infty}^{-}  &  =\frac{1}{2}-\frac{1}{2}\sqrt{1+4b}%
\end{align*}
Here $b=\frac{\operatorname{lcoef}(s)}{\operatorname{lcoeff}(t)}$ where $r=\frac{s}{t}$. $\operatorname{lcoef}(s)$ is
the leading coefficient of $s$ and similarly, $\operatorname{lcoef}(t)$ is the leading coefficient of $t$.
\item If $\mathcal{O}(\infty)>2$ then
\begin{align*}
\left[  \sqrt{r}\right]  _{\infty}  &  =0\\
\alpha_{\infty}^{+}  &  =0\\
\alpha_{\infty}^{-}  &  =1
\end{align*}
\end{enumerate}

\subsubsection{step 2}
Using quantities calculated in step $1$, the algorithm now
searches for a non-negative integer $d$ using
\begin{align*} 
d&=\alpha_{\infty}^{\pm}-\sum_{c\in \Gamma}\alpha_{c}^{\pm}
\end{align*} 
If non-negative $d$ is found, a candidate $\omega_{d}$ is calculated using%
\begin{align*} 
\omega_{d}&=\sum_{c\in \Gamma}\left(  (\pm) \left[  \sqrt{r}\right]
_{c}+\frac{\alpha_{c}^{\pm}}{x-c}\right)  +  (\pm)  \left[\sqrt{r}\right]_{\infty}
\end{align*} 

If no non-negative integer $d$ could be found, then no Liouvillian solution 
exists using this case. Case two or three are tried next if these are available.
\subsubsection{step 3}
In this step the algorithm finds polynomial $p(x)=a_0+a_1 x+ a_2 x^2 + \dots + x^d$ of degree $d$. This is done
by solving for the coefficients $a_i$ from
\begin{align*} 
p^{\prime\prime}+2\omega p^{\prime}+\left(  \omega^{\prime}+\omega^{2}-r\right)  p&=0 \tag{7} 
\end{align*} 
Where $\omega$ is from the second step above and $r$ is from $z''=rz$. 

For an example, if $d=2$, then $p\left(  x\right)  =x^{2}+a_1 x + a_0$ is 
substituted in (3) and $a_0,a_1$ are solved for. If solution exists,
then the solution to $z''=r z$ will be 
\begin{align*}
z&=p(x) e^{\int\omega dx}
\end{align*}
If the degree $d=1$  then $p\left(  x\right)=x+a_0$ and the same process is applied. If the degree $d=0$, then 
$p\left(x\right)=1$. 

The first basis solution to the original ode is now be found from
\begin{align*}
y_1&=z e^{-\frac{1}{2}  \int a \, dx}
\end{align*}
And the second basis solution using reduction of order formula is
\begin{align*}
y_2&=y_1 \int{   \frac{  e^{-\int a \, dx}  }{y_{1}^{2}}          \, dx} 
\end{align*}
Hence the general solution to the original ode is
\begin{align*}
y(x) = c_1 y_1 + c_2 y_2
\end{align*}
This completes the full algorithm for case 1. The part that needs most care is 
in finding $\left\{\left[\sqrt{r}\right]  _{c},\alpha_{c}^{\pm},\left[  \sqrt{r}\right]  _{\infty
},\alpha_{\infty}^{\pm}\right\}$. Once these are calculated, the rest of the algorithm is
much more direct.
\subsubsection{Algorithm flow chart for case one} 
\begin{figure}[H]
\centering
\fbox{\includegraphics[width=.9\textwidth]{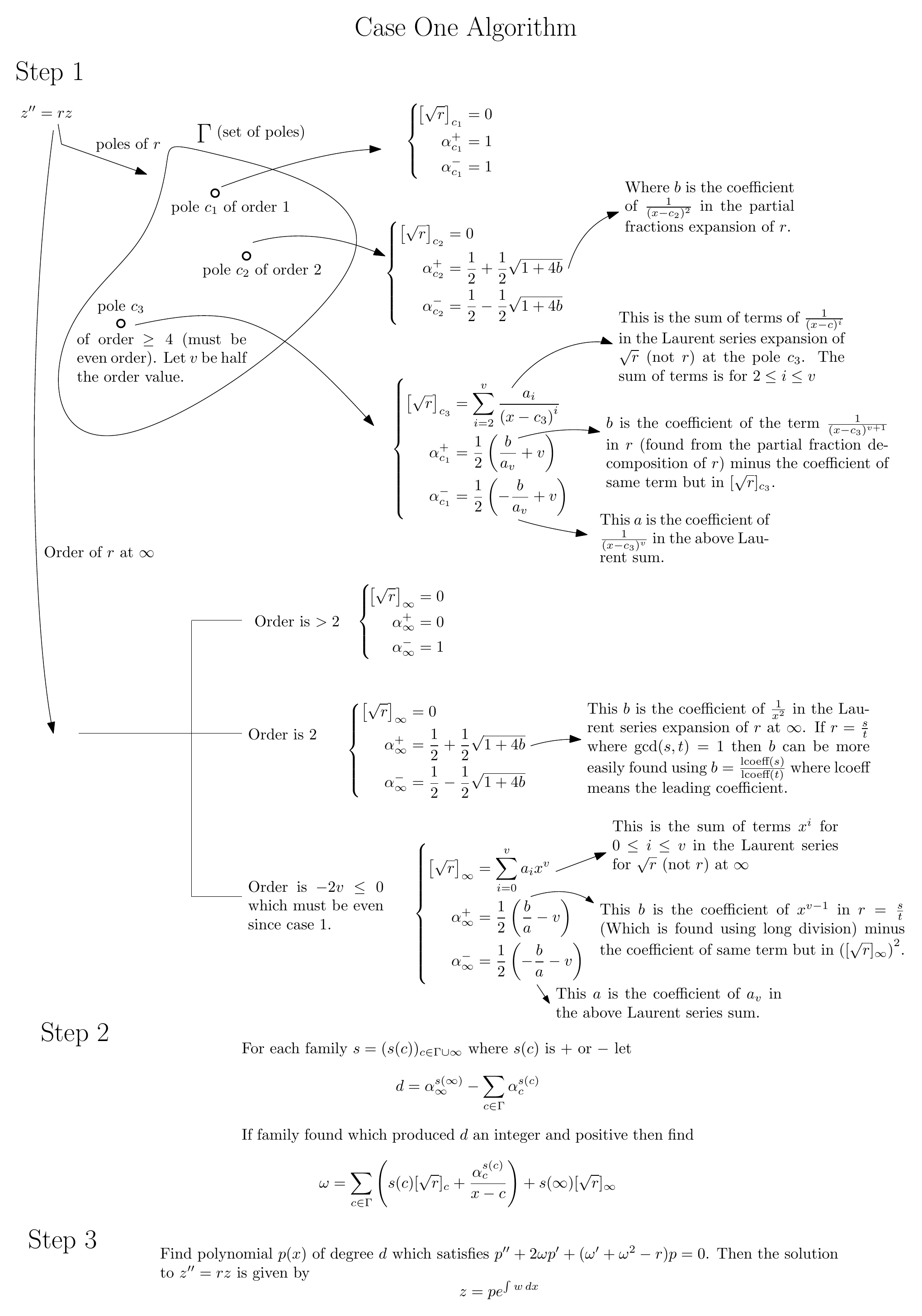}}
\caption{Case 1 Kovacic algorithm}
\end{figure}

\subsection{Case two}
\subsubsection{step 1}
Assuming that the necessary conditions for case two are satisfied 
and $z''=r z, r=\frac{s}{t}$. Let $\Gamma$  be the set of all poles 
of $r$. For each pole $c$ in this set, $E_c$ is found as follows
\begin{enumerate}
\item If the pole $c$ has order $1$ then $E_c=\{4\}$.
\item If the pole $c$ is of order $2$ then $E_c=\{2, 2+ 2\sqrt{1+ 4b}, 2-2 \sqrt{1+ 4b}\}$ where
$b$ is the coefficient of $\frac{1}{(x-c)^2}$ in the partial fraction decomposition of $r$. In
the above set $E_c$, only integer values are kept.
\item If the pole $c$ is of order $v>2$ then $E_c=\{v\}$
\end{enumerate}
The next step is to determine $E_{\infty}$.
\begin{enumerate}
\item If $\mathcal{O}(\infty)>2$ then  $E_{\infty}=\{0,2,4\}$
\item If $\mathcal{O}(\infty)=2$ then $E_{\infty}=\{2, 2+ 2\sqrt{1+ 4b}, 2-2 \sqrt{1+ 4b}\}$ where 
$b=\frac{\operatorname{lcoef}(s)}{\operatorname{lcoeff}(t)}$ where $r=\frac{s}{t}$. $\operatorname{lcoef}(s)$ is
the leading coefficient of $s$ and similarly $\operatorname{lcoef}(t)$ is the leading coefficient of $t$.
In the above set $E_{\infty}$ only integer values are kept.
\item If $\mathcal{O}(\infty)<2$ then $E_{\infty}=\mathcal{O}(\infty)$.
\end{enumerate}
\subsubsection{step 2} 
Using quantities calculated in step $1$, the algorithm now searches for a non-negative integer $d$ using
\begin{align*} 
d&=\frac{1}{2} \left( e_{\infty} - \sum_{c\in \Gamma} e_c \right) 
\end{align*} 
Where in the above $e_c \in E_c$, $e_\infty \in E_\infty$ found in step $1$. If non-negative
$d$ is found, then 
\begin{align*} 
\theta &= \frac{1}{2} \sum_{c\in \Gamma} \frac{e_c}{x-c}
\end{align*} 
If no non-negative integer $d$ could be found, then no Liouvillian solution 
exists using this case. Case three is tried next if it is available.
\subsubsection{step 3}  
In this step the algorithm determines a polynomial $p(x)=a_0+a_1 x+ a_2 x^2 + \dots + x^d$ of degree $d$. 
This is done by solving for the coefficients $a_i$ from 
\begin{align*} 
p'''+3 \theta p'' + \left(3 \theta^2 + 3 \theta' - 4 r \right) p' + \left(\theta''+ 3 \theta \theta' + \theta^3 - 4 r \theta -2 r' \right)p=0   \tag{1} 
\end{align*}  
Where $\theta$ was found in step $2$ and $r$ is from $z''=rz$. If $p(x)$ can be found that satisfies (1) then
\begin{align*} 
\phi = \theta + \frac{p'}{p}  \tag{2} 
\end{align*} 
$\omega$ is then solved for from
\begin{align*} 
\omega^2 - \phi \omega + \left(\frac{1}{2} \phi' +\frac{1}{2} \phi^2 -r\right) &=0 \tag{3} 
\end{align*} 
If solution $\omega$ to (3) can be found, then the solution to $z''=rz$ is given by
\begin{align*} 
z = e^{\int{ \omega \,dx}}
\end{align*} 
This completes the full algorithm for case two.  The general solution to the original ode
is now determined as outlined at the end of case one above.
\subsubsection{Algorithm flow chart for case two}
\begin{figure}[H]
\centering
\fbox{\includegraphics[width=.9\textwidth]{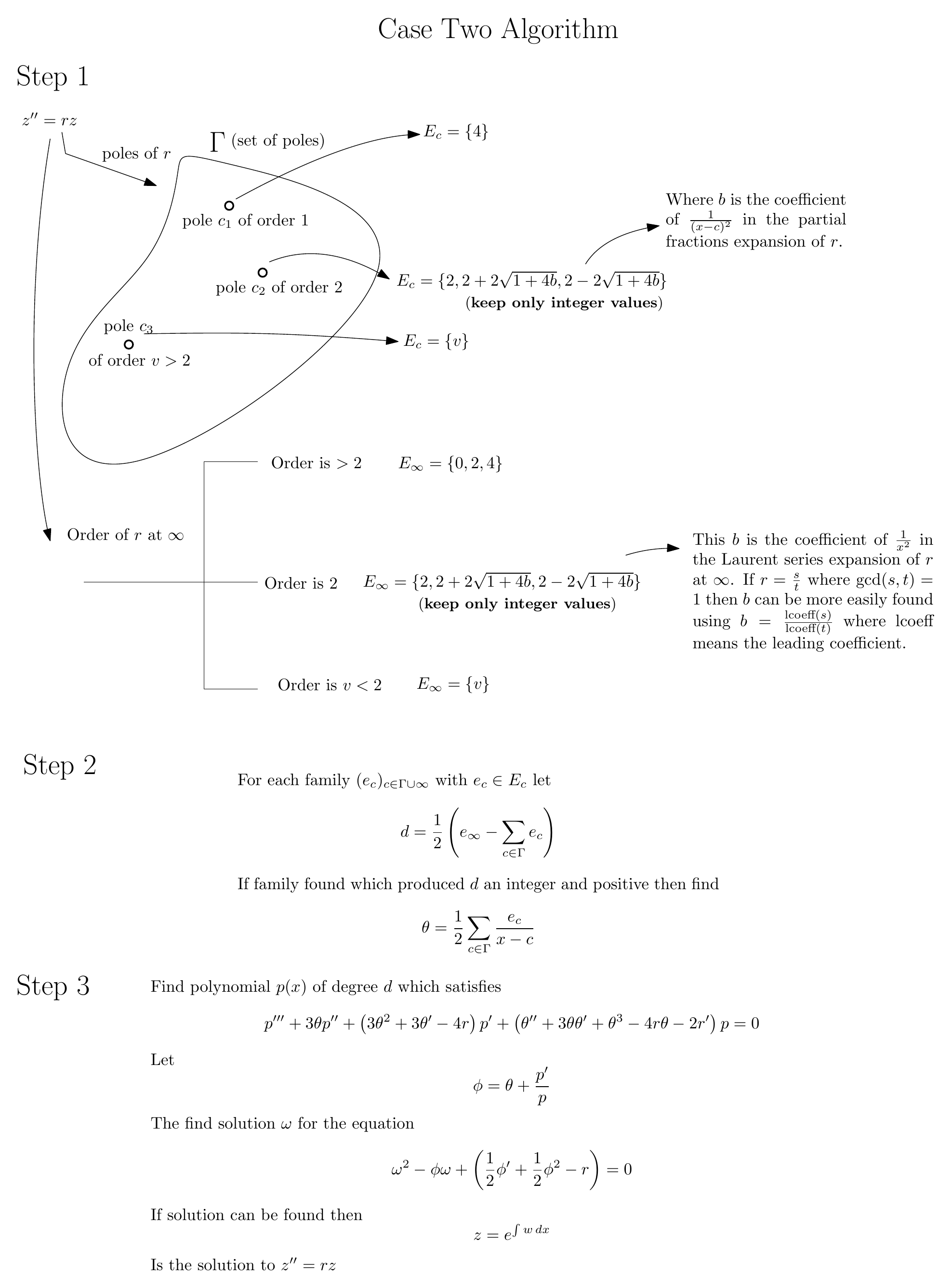}}
\caption{Case 2 Kovacic algorithm}
\end{figure}
\subsection{Case three}
\subsubsection{step 1}
Assuming the necessary conditions for case three are satisfied  
and $z''=r z, r=\frac{s}{t}$. Let $\Gamma$  be the set of all poles 
of $r$. Recall that case three can have either a pole of order 1 or order 2 only. For 
each pole $c$ in this set, $E_c$ is found as follows
\begin{enumerate}
\item If the pole $c$ has order $1$ then $E_c=\{12\}$.
\item If the pole $c$ is of order $2$ then 
\begin{alignat*}{2}
E_c &=\left\{ 6 + \frac{12 k}{n} \sqrt{1+ 4 b} \right\} \qquad \text{for} \quad  k &&=-\frac{n}{2}\cdots \frac{n}{2}   \tag{1} 
\end{alignat*} 
Where $k$ is incremented by $1$ each time,  and $n$ is any of $\{4,6,12\}$ and $b$ is the 
coefficient of $\frac{1}{(x-c)^2}$ in the partial fraction decomposition of $r$. In
the above set $E_c$, only integer values are kept. For an example, when $n=4$ then $k=\{-2,-1,0,1,2\}$ and
$E_c =\{ 6 -6 \sqrt{1+ 4 b},6 -3 \sqrt{1+ 4 b},6,6 + 3 \sqrt{1+ 4 b},6 + 6 \sqrt{1+ 4 b}\}$ and 
similarly for $n=6$ and $n=12$.
\end{enumerate}
The next step determines $E_{\infty}$. This is found using same formula as (1) but $b$ is
calculated differently using
$b=\frac{\operatorname{lcoef}(s)}{\operatorname{lcoeff}(t)}$ where $r=\frac{s}{t}$. $\operatorname{lcoef}(s)$ is
the leading coefficient of $s$ and $\operatorname{lcoef}(t)$ is the leading coefficient of $t$.
\subsubsection{step 2}
Using quantities calculated in step $1$, the algorithm now searches for a non-negative integer $d$ using
\begin{align*} 
d&=\frac{n}{12} \left( e_{\infty} - \sum_{c\in \Gamma} e_c \right) 
\end{align*} 
Where in the above $e_c \in E_c$, $e_\infty \in E_\infty$ 
$n$ is any of $\{4,6,12\}$ values. If non-negative $d$ is found, then 
\begin{align*} 
\theta &= \frac{n}{12} \sum_{c\in \Gamma} \frac{e_c}{x-c}
\end{align*} 
The sum above is over all families of $\{e_\infty,e_c\}$ which generated the non-negative integer $d$.
Next define
\begin{align*} 
S &= \prod_{c\in \Gamma} (x-c)
\end{align*}  
The product above is over families of $\{e_\infty,e_c\}$ which generated the non-negative integer $d$.
If no non-negative integer $d$ is found, then no Liouvillian solution exists.
\subsubsection{step 3} 
In this step the algorithm determines a polynomial $p(x)=a_0+a_1 x+ a_2 x^2 + \dots + x^d$ of degree $d$.
Define set of polynomials $\{P_n,P_{n-1},\cdots,P_{-1}$ where
\begin{align*}  
 P_n &= -p(x) && &&\\
 P_{i-1} &= -S P_{i}' +\left( (n-i) S' - S \theta\right)P_i -(n-i)(i+1) S^2 r P_{i+1}   \qquad  i=n \cdots 0
\end{align*}      
The last  polynomial $P_{-1}(x)$ is used to solve for the coefficients $a_i$ using
\begin{align*} 
 P_{-1}(x) &= 0 \tag{2}  
\end{align*}
In Maple this is done using the \verb|solve| command with the \verb|identity| option.
If it is possible to find coefficients $a_i$ such that (2) is satisfied, then define the
equation
\begin{align*}  
  \sum_{i=0}^{n} \frac{S^i P_{i}(x)}{(n-i)!} \omega^i   = 0
\end{align*}
$\omega$ is solved for from the above equation. If solution $\omega$ is found then the solution
to $z''=r z$ will be 
\begin{align*} 
 z &= e^{\int \omega \,dx}
\end{align*}
This completes the full algorithm for case three.   The general solution to the original ode
can now be determined as outlined at the end of case one above.
\subsubsection{Algorithm flow chart for case three} 
\begin{figure}[H]
\centering
\fbox{\includegraphics[width=.9\textwidth]{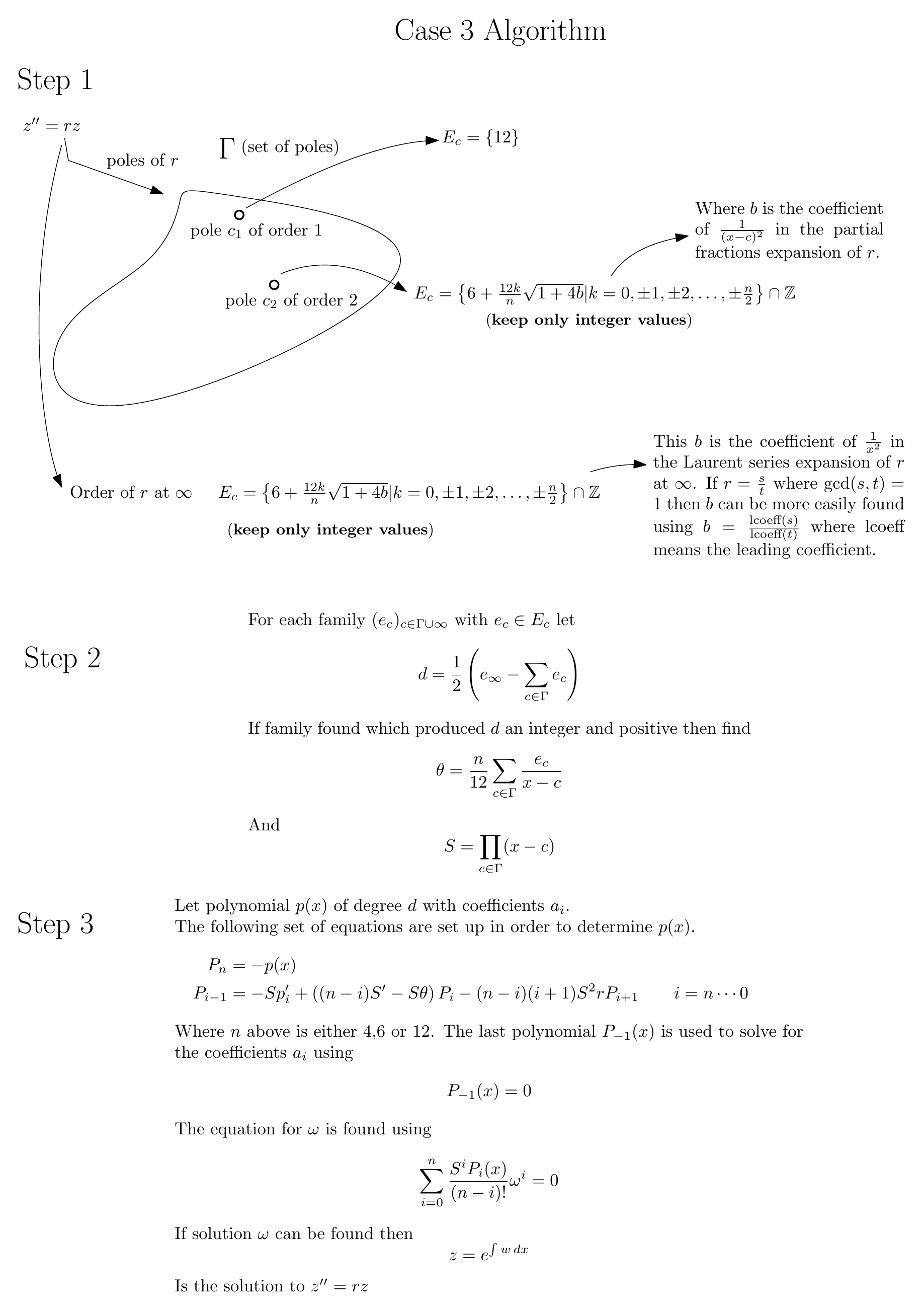}}
\caption{Case 3 Kovacic algorithm}
\end{figure} 
\clearpage

\subsection{Statistics and discussion of results obtained using Kovacic algorithm}
This gives summary of results obtained using testsuite of $3000$ differential 
equations, all of which were selected as linear with rational 
coefficients as functions of $x$ that can be solved using this algorithm.

The ode's used in the testsuite were collected by the author 
and stored in sql database. These were collected from a number of 
standard textbooks and other references such  as 
``Differential Equations. E. Kamke. 3th edition. Chelsea.'' and 
``Ordinary Differential Equations And Their Solutions. Murphy, George Moseley. Dover. 2011''.

All the ode's were successfully solved using the Kovacic algorithm as 
implemented here and each solution was verified using Maple odetest. 

The following diagram shows the percentage of ode's solved using each case.

\begin{figure}[H]
\centering
\fbox{\includegraphics[width=.7\textwidth]{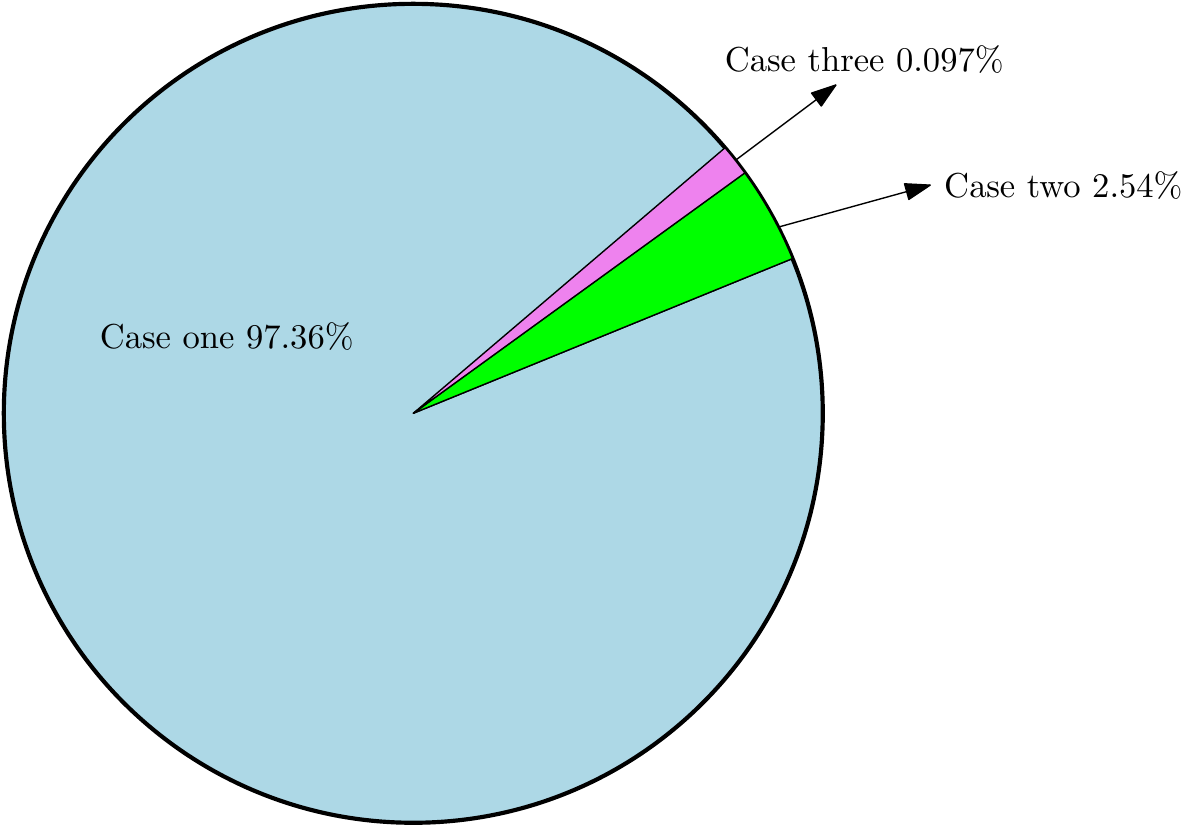}}
\caption{Kovacic cases distributions}
\end{figure} 

Case $3$ was required for solving only $3$ odes. It used $n=4$ for all $3$ ode's. $n=6$ and $n=12$ 
were not reached or required to try. Recall that $n$ for case $3$ is the degree of 
the polynomial in $\omega$ used to solve for in order to find 
the $z$ solution from $z=e^{\int \omega \,dx}$. 

This result shows that case $1$ and $2$ combined is all what is needed to solve $99.9$\% of 
ode's used in practice. Larger collection of ode's than the $3000$ used could produce
different results, but the overall trend is that case $3$ is rarely needed in practice
and within case $3$, $n=6$ and $n=12$ are even less likely to be required.

When forcing the algorithm to use case $3$ and only use $n=12$, 
this resulted in a very long computation time on some ode's. For an example, 
using ode $y''+xy'+y=0$ which satisfies all three cases, and asking the solver 
to use case $3$ and $n=12$, it was found that it required $p(x)$ of degree $d=24$ 
in order to find $\omega$  of degree $12$ that can be solved. The total number of trials 
in step 3 of case three to find such solution was found to be $2367$. This took over 
30 minutes to complete.  

In comparison, the same ode was solved using case one in less than one second 
giving the same solution on the same computer.

The testsuite also calculates the distribution of cases which has its necessary conditions
satisfied for each ode. Recall that having the necessary conditions for a case satisfied
does not mean a solution would be found using that case.  The following
bar chart shows the percentages of the $3000$ ode's that satisfied the 
necessary conditions each case. This chart shows that many ode's satisfy the
conditions for more than one case at the same time.

\begin{figure}[H]
\centering 
\fbox{\includegraphics[width=.7\textwidth]{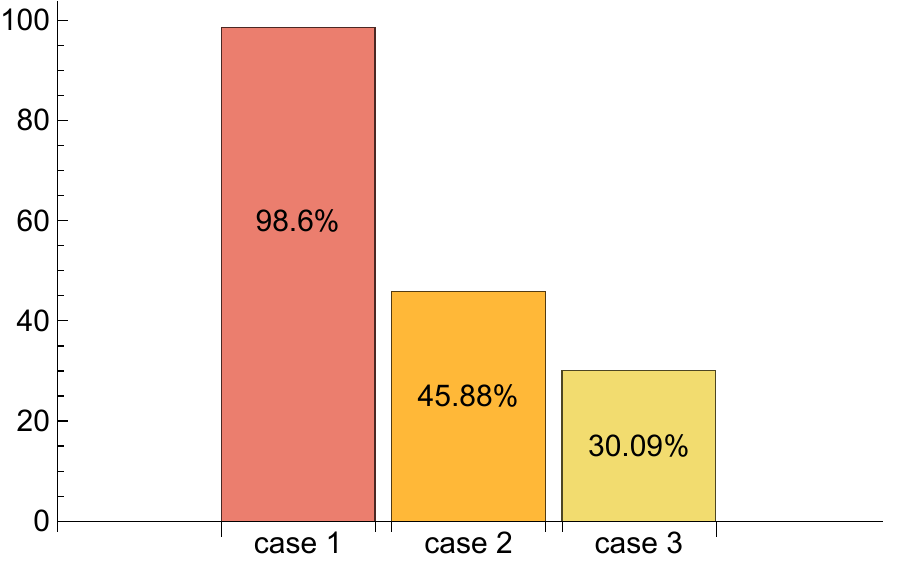}}
\caption{Percentage of ode's that satisfy  each Kovacic case necessary conditions}
\end{figure} 
\clearpage
\section{Worked example for each case}
\subsection{case one}

\subsubsection{Example 1}
Given the ode
\begin{align*}
(2 x + 1)y''-2 y'- (2x +3)y=0
\end{align*}
Converting it $y''+ a y' + b y=0$ gives
\begin{align*}
y''-\frac{2}{2 x + 1} y'-\frac{2x +3}{2 x + 1} y=0
\end{align*}
Where $a=-\frac{2}{2 x + 1}, b=-\frac{2x +3}{2 x + 1}$. 
Applying the transformation $z=y e^{\frac{1}{2} \int{a\,dx}}$ gives $z''=r z$ where
$r = \frac{1}{4} a^2 + \frac{1}{2} a' - b$. This results in
\begin{align*}
r &=\frac{s}{t}\\
  &=\frac{4x^2+8x+6}{(2x+1)^2}=\frac{s}{t}
\end{align*}
There is one pole at $x=-\frac{1}{2}$, hence $\Gamma=\{-\frac{1}{2}\}$. The order is $2$ and $\mathcal{O}(\infty)=\deg(t)-\deg(s)=0$. 
Table~\ref{tab:first} shows that the necessary conditions for case one and two are both satisfied. 
This is solved first using case one. Since the order of the pole is $2$, then
\begin{align*}\tag{1}
\left[  \sqrt{r}\right]_{c}  &  =0\\
\alpha_{c}^{+}  &  =\frac{1}{2}+\frac{1}{2}\sqrt{1+4b}\\
\alpha_{c}^{-}  &  =\frac{1}{2}-\frac{1}{2}\sqrt{1+4b}
\end{align*}
Where $b$ is the coefficient of 
$\frac{1}{(x-c)^{2}}=\frac{1}{(x+\frac{1}{2})^{2}}$ in the partial fraction decomposition of $r$
which is $r=1+\frac{3}{4}\frac{1}{(x+\frac{1}{2})^2}+\frac{1}{x+\frac{1}{2}}$. 
Therefore $b=\frac{3}{4}$ and the above becomes
\begin{align*}
\left[  \sqrt{r}\right]_{c}  &  =0\\
\alpha_{c}^{+}  &  =\frac{1}{2}+\frac{1}{2}\sqrt{1+4\left(\frac{3}{4}\right)} = \frac{3}{2}\\
\alpha_{c}^{-}  &  =\frac{1}{2}-\frac{1}{2}\sqrt{1+4\left(\frac{3}{4}\right)} = -\frac{1}{2}
\end{align*}
Since $\mathcal{O}(\infty)= 0$ then $v =0$ and $\left[  \sqrt{r}\right]  _{\infty}$ is the sum 
of all terms $x^i$ for $0\leq i\leq v$ in the Laurent series expansion of $\sqrt{r}$ at $\infty$
which is found as follows. Since  $\sqrt{r} =\sqrt{\frac{4x^2+8x+6}{(2x+1)^2}}$ then setting $x=\frac{1}{y}$
gives $\sqrt{r(y)} =\sqrt{\frac{4 \left(\frac{1}{y}\right)^2+8\frac{1}{y}+6}{\left(2\frac{1}{y}+1\right)^2}}$ and
since $v=0$ then the constant term is $\lim_{y\rightarrow 0} \sqrt{r(y)}= 1$. Therefore $\left[  \sqrt{r(x)}\right]  _{\infty}=1$. 
Hence $a=1$.  

$b$ is the coefficient of
$x^{v-1}=\frac{1}{x}$ in $r$ minus the coefficient of $\frac{1}{x}$ in  $\left(\left[  \sqrt{r}\right]  _{\infty}\right)^2=1$ which is
zero since there is no term $\frac{1}{x}$. Because $v=0$, long division is used is used find 
the coefficient of $\frac{1}{x}$ in $r$.
\begin{align*} 
r &= \frac{s}{t}\\
  &= \frac{4 x^2 + 8 x + 6}{4 x^2 + 4 x +1}\\
  &= Q + \frac{R}{t}\\
  &= 1 + \frac{4 x + 5}{4 x^2+4 x+1}
\end{align*} 
The coefficient of  $\frac{1}{x}$  in $r$ is the leading coefficient in $R$ minus 
the leading coefficient in $t$ which gives $\frac{4}{4}=1$.
Therefore $b=1-0=1$. This results in
\begin{alignat*}{3}
\left[  \sqrt{r}\right]  _{\infty}&=1\\
\alpha_{\infty}^{+}  &  =\frac{1}{2}\left(  \frac{b}{a_v}-v\right) &&= \frac{1}{2}\left(  \frac{1}{1}-0\right) &&= \frac{1}{2}\\
\alpha_{\infty}^{-}  &  =\frac{1}{2}\left(  -\frac{b}{a_v}-v\right) &&= \frac{1}{2}\left(  -\frac{1}{1}-0\right) &&= -\frac{1}{2}
\end{alignat*}
The above completes step 1 for case one. Step $2$ searches for a non-negative integer $d$ using
\begin{align*} 
d&=\alpha_{\infty}^{\pm}-\sum_{c\in \Gamma}\alpha_{c}^{\pm} \tag{2}
\end{align*} 
Where the above is carried over all possible combinations resulting in the following
4 possibilities (in this example, there is only one pole, hence the sum contains only one term) and
the number of possible combinations is therefore $2^2=4$
\begin{alignat*}{3} 
d &= \alpha_{\infty}^{+} -  \alpha_{c}^{+} &&= \frac{1}{2} -   \left( \frac{3}{2}\right) &&= -1\\
d &= \alpha_{\infty}^{+} -  \alpha_{c}^{-} &&= \frac{1}{2} -   \left(- \frac{1}{2}\right) &&= 1 \\
d &= \alpha_{\infty}^{-} -  \alpha_{c}^{+} &&= -\frac{1}{2} -  \left( \frac{3}{2}\right) &&= -2 \\
d &= \alpha_{\infty}^{-} -  \alpha_{c}^{-} &&= -\frac{1}{2} -  \left(- \frac{1}{2}\right) &&= 0
\end{alignat*} 
The above shows there are two possible $d$ values to use. $d=1$ or $d=0$. Each is tried until one
produces a solution or all fail to do so.
For each valid $d$ found an  $\omega$ is found using
\begin{align*} 
\omega&=\sum_{c}\left(  (\pm) \left[  \sqrt{r}\right]_{c}+\frac{\alpha_{c}^{\pm}}{x-c}\right)  +  (\pm)  \left[\sqrt{r}\right]_{\infty}
\end{align*} 
But $\left[  \sqrt{r}\right]_{c}=0$ in this example, hence the above simplifies to 
\begin{align*} 
\omega&=\sum_{c}\left( \frac{\alpha_{c}^{\pm}}{x-c}\right)  +  (\pm)  \left[\sqrt{r}\right]_{\infty}
\end{align*}
Since there is one pole, then the candidate  $\omega$ to try are the following
\begin{align*} 
\omega&= \frac{\alpha_{c}^{+}}{x-c} +  (+1)  \left[\sqrt{r}\right]_{\infty}\\
\omega&= \frac{\alpha_{c}^{+}}{x-c} +  (-1)  \left[\sqrt{r}\right]_{\infty}\\
\omega&= \frac{\alpha_{c}^{-}}{x-c} +  (+1)  \left[\sqrt{r}\right]_{\infty}\\
\omega&= \frac{\alpha_{c}^{-}}{x-c} +  (-1)  \left[\sqrt{r}\right]_{\infty}
\end{align*}
Substituting the known values found in step 1 into the above gives
\begin{align*} 
\omega&= \frac{ \frac{3}{2} } {x+\frac{1}{2}} +  (+1)  (1)  = \frac{6+2 x}{2 x +3} \\
\omega&= \frac{ \frac{3}{2} } {x+\frac{1}{2}} +  (-1)  (1)  = -\frac{2 x}{2 x +3}\\
\omega&= \frac{ -\frac{1}{2} } {x+\frac{1}{2}} +  (+1)  (1) = \frac{2 x}{2 x +1}\\
\omega&= \frac{ -\frac{1}{2} } {x+\frac{1}{2}} +  (-1) (1) = -\frac{2 \left(1+x \right)}{2 x +1}
\end{align*}
So there are two possible $d$ values to try, and for each, there are $4$ possible $w(x)$, 
which gives $8$ possible tries. This completes step 2.  For each trial, step 3 is now invoked. 

Starting with $d=0$ and using $\omega=\frac{2 x}{2 x +1}$, and since the degree is $d=0$ then 
$p(x)=1$. This polynomial is now checked to see if it satisfies
\begin{align*}
 p''+2 \omega p' + (\omega'+\omega^2-r) p &= 0  \tag{3}\\
  (\omega'+\omega^2-r) p &= 0 \\
 \frac{d}{dx} \left(\frac{2 x}{2 x +1}\right)   +\left(\frac{2 x}{2 x +1}\right)^2- \frac{4x^2+8x+6}{(2x+1)^2}   &=0\\
-\frac{4}{2 x +1} &=0
\end{align*}
Since the left side is not identically zero, then this candidate $\omega$ has failed. Carrying out 
this process for the other 3 possible $\omega$ values shows that non are satisfied as well. $d=1$ is now tried. 
This implies the polynomial is $p(x)=a_0 + x$. The coefficient $a_0$
needs to be determined such that $p''+2 \omega p' + (\omega'+\omega^2-r) p= 0$ is satisfied. Starting with 
$\omega= \frac{2 x}{2 x +1}$ gives
\begin{align*}
 p''+2 \omega p' + (\omega'+\omega^2-r) p &= 0 \\
 2 \omega p' + (\omega'+\omega^2-r) p &= 0
\end{align*}
Substituting $p=a_0+x$ and $\omega= \frac{2 x}{2 x +1}$ and $r= \frac{4x^2+8x+6}{(2x+1)^2}$ into 
the above and simplifying gives $-\frac{4 a_0}{2 x +1}=0$. This implies that (3) can be satisfied for $a_0=0$. 
Therefore the polynomial of degree one is found and given by
\begin{align*}
p(x)&=x
\end{align*}
Therefore the solution to $z''=rz$ is 
\begin{align*}
z&=p e^{\int \omega\, dx}\\
  &=x e^{\int \frac{2 x}{2 x +1} \, dx}\\
  &=x e^{x -\frac{\ln \left(2 x +1\right)}{2}}\\
  &=\frac{x e^x}{\sqrt{2 x + 1}}
\end{align*}
Given this solution for $z(x)$, the first basis solution of the original ode in $y$ is found
using the inverse of the original transformation used to generate the $z$ ode which is
$z = y e^{\frac{1}{2}\int a\,dx}$, therefore 
\begin{align*}
y_1 &= z e^{-\frac{1}{2}\int a\,dx}\\
    &= \frac{x e^x}{\sqrt{2 x + 1}} e^{-\frac{1}{2}\int -\frac{2}{2 x + 1}   \,dx}\\
    &= \frac{x e^x}{\sqrt{2 x + 1}} e^{\frac{\ln(2x+1}{2}}\\
    &= \frac{x e^x}{\sqrt{2 x + 1}} \sqrt{2 x + 1}\\
    &= x e^x
\end{align*}
The second basis solution is found using reduction of order
\begin{align*}
y_2 &= y_1 \int{  \frac{   e^{\int{-a \,dx}}\,dx } {y_{1}^{2}}  \, dx } \\
    &= x e^x \int{  \frac{   e^{\int{- \frac{-2}{2 x + 1} \,dx}}\,dx } {(x e^x)^{2}}  \, dx } \\
    &= x e^x \int{  \frac{   e^{\ln(2 x + 1)}} {(x e^x)^{2}}  \, dx } \\
    &= -e^{-x}
\end{align*}
Therefore the general solution to the original ode 
$(2 x + 1)y''-2 y'- (2x +3)y=0$ is
\begin{align*}
y(x) &= c_1 y_1 + x_2 y_2 \\
y(x) &= c_1 xe^x - c_2 e^{-x}
\end{align*}
This completes the solution.
\subsubsection{Example 2}
Given the ode  
\begin{align*}
x^2 \left(x^2-2 x +1\right) y''-x (3+x) y'+(4+x) y=0
\end{align*}
Converting it $y''+ a y' + b y=0$ gives
\begin{align*}
y''- \frac{x(3+x)}{x^2 \left(x^2-2 x +1\right)} y'+ \frac{4+x}{x^2 \left(x^2-2 x +1\right)} y=0
\end{align*}
Where $a=- \frac{x (3+x) }{x^2 \left(x^2-2 x +1\right)}, b=\frac{4+x}{x^2 \left(x^2-2 x +1\right)}$. 
Applying the transformation $z=y e^{\frac{1}{2} \int{a\,dx}}$ gives $z''=r z$ where
$r = \frac{1}{4} a^2 + \frac{1}{2} a' - b$. This results in
\begin{align*}
r &= \frac{s}{t} \\
  &= \frac{7 x^{2}+10 x -1}{4 x^{2} \left(x -1\right)^{4}}
\end{align*}
There is one pole at $x=0$ of order $2$ and pole at $x=1$ of order $4$, hence $\Gamma=\{0,1\}$, 
and $\mathcal{O}(\infty)=\deg(t)-\deg(s)=6-2=4$. Table~\ref{tab:first} shows that 
the necessary conditions for case one and two are both satisfied. 
For the pole  at $0$ and since its order is $2$ then
\begin{align*}\tag{1}
\left[  \sqrt{r}\right]_{0}  &  =0\\
\alpha_{0}^{+}  &  =\frac{1}{2}+\frac{1}{2}\sqrt{1+4b}\\
\alpha_{0}^{-}  &  =\frac{1}{2}-\frac{1}{2}\sqrt{1+4b}
\end{align*}
Where $b$ is the coefficient of 
$\frac{1}{(x-0)^{2}}=\frac{1}{x^{2}}$ in the partial fraction decomposition of $r$
which is 
\begin{align*}
r &= \frac{4}{\left(x -1\right)^{4}}-\frac{2}{\left(x -1\right)^{3}}-\frac{1}{4 x^{2}}-\frac{3}{2 \left(x -1\right)}+\frac{3}{2 x}+\frac{7}{4 \left(x -1\right)^{2}} \tag{2} 
\end{align*}
The above shows that $b=-\frac{1}{4}$. Equation (1) becomes
becomes
\begin{align*}
\left[  \sqrt{r}\right]_{0}  &  =0\\
\alpha_{0}^{+}  &  =\frac{1}{2}+\frac{1}{2}\sqrt{1-4\frac{1}{4}} = \frac{1}{2}\\
\alpha_{0}^{-}  &  =\frac{1}{2}-\frac{1}{2}\sqrt{1-4\frac{1}{4}} = \frac{1}{2}
\end{align*}
For the second pole at $x=1$, since its order is $4$, then $2v=4$ or $v=2$. Therefore the corresponding
$\left[  \sqrt{r}\right]_{1}$ is the sum of all terms involving $\frac{1}{ (x-1)^i}$ 
for $2\leq i \leq v$ or $2\leq i \leq 2$ in the Laurent series expansion of 
$\sqrt{r}$ (not $r$) around this pole. This results in
\begin{align*} 
[\sqrt{r}]_{1} &= \sum_{i=2}^{2} \frac{a_i}{(x-1)^i}\\
               &= \frac{a_2}{(x-1)^2} \tag{3}
\end{align*} 
$a_2$ is found using 
\begin{align*} 
a_2 &= \lim_{x\rightarrow 1} (x-1)^2 \sqrt{r}\\
    &= \lim_{x\rightarrow 1} (x-1)^2 \sqrt{   \frac{7 x^{2}+10 x -1}{4 x^{2} \left(x -1\right)^{4}}  }\\
    &= 2
\end{align*} 
Therefore
\begin{alignat*}{3}
[\sqrt r]_{1}  &=  \frac{2}{ (x-1)^2} \\ 
\alpha_{1}^{+}   &=  \frac{1}{2} \left( \frac{b}{a} + v \right)  &&= \frac{1}{2} \left( \frac{b}{2} + 2 \right) \\ \tag{4}
\alpha_{1}^{-}    &=  \frac{1}{2} \left(- \frac{b}{a} + v \right) &&= \frac{1}{2} \left(- \frac{b}{2} + 2 \right)
\end{alignat*}
What remains is to determine $b$. This is the coefficient of $\frac{1}{ (x-c)^{v+1}}=\frac{1}{ (x-1)^{3}}$ in 
the partial fraction decomposition of $r$ which from (2) is $-2$  minus the coefficient of same term in $[\sqrt r]_{1}$
which from (3) is zero. Therefore $b=-2-0=-2$. (4) now becomes
\begin{alignat*}{3}
[\sqrt r]_{1}  &=  \frac{2}{ (x-1)^2} \\ 
\alpha_{1}^{+}   &=  \frac{1}{2} \left( \frac{b}{a} + v \right)  &&= \frac{1}{2} \left( \frac{-2}{2} + 2 \right) &&=\frac{1}{2} \\ \tag{5}
\alpha_{1}^{-}    &=  \frac{1}{2} \left(- \frac{b}{a} + v \right) &&= \frac{1}{2} \left(- \frac{-2}{2} + 2 \right) &&=\frac{3}{2}
\end{alignat*}
The above completes finding $[\sqrt r]_{c},\alpha_{c}^{+},\alpha_{c}^{+}$ for all poles in the set $\Gamma$.

Since the order of $r$ at $\infty$ is $4 > 2$ then
\begin{alignat*}{2}
       [\sqrt r]_\infty     &=  0 \\ 
      \alpha_{\infty}^{+}  &=  0 \\ 
       \alpha_{\infty}^{-}  &=  1
\end{alignat*}
\begin{center}
\begin{minipage}{\textwidth}
This completes the first step of the solution. The following tables summarizes the findings so far
\begin{table}[H]
\centering
\begin{tabular}[c]{|c|c|c|c|c|}\hline
pole $c$ location & pole order & $[\sqrt r]_c$ & $\alpha_c^{+}$ & $\alpha_c^{-}$ \\ \hline
 $0$ & $2$ & $0$ &  ${\frac{1}{2}}$ &  ${\frac{1}{2}}$ \\ \hline
 $1$ & $4$ & $\frac{2}{\left(x -1\right)^{2}}$ &  ${\frac{1}{2}}$ &  ${\frac{3}{2}}$ \\ \hline
\end{tabular}
\caption{First step, case one. $\Gamma$ set information}\label{tab:second}
\end{table}

\begin{table}[H]
\centering
\begin{tabular}[c]{|c|c|c|c|}\hline
Order of $r$ at $\infty$ &  $[\sqrt r]_\infty $ & $\alpha_\infty^{+}$ & $\alpha_\infty^{-}$ \\ \hline
$4$ & $0$ & $0$ & $1$\\ \hline 
\end{tabular}
\caption{First step, case one.  $\mathcal{O}(\infty)$ information}\label{tab:third}
\end{table}
\end{minipage}
\end{center}

This completes step 1 for case one. Step 2 searches for non-negative integer $d$ using
\begin{align*} 
d&=\alpha_{\infty}^{\pm}-\sum_{c\in \Gamma}\alpha_{c}^{\pm}
\end{align*} 
Where the above is carried over all possible combinations resulting in the following
8 possibilities (in this example, there are two poles, hence the sum contains two term) and
the number of possible combinations is therefore $2^3=8$
\begin{alignat*}{3} 
d &= \alpha_{\infty}^{+} - \left(  \alpha_{0}^{+} + \alpha_{1}^{+}\right) &&= 0 -\left( \frac{1}{2} + \frac{1}{2}    \right)  &&= -1  \\
d &= \alpha_{\infty}^{+} - \left(  \alpha_{0}^{+} + \alpha_{1}^{-}\right) &&= 0 -\left( \frac{1}{2} + \frac{3}{2}    \right)  &&= -2\\
d &= \alpha_{\infty}^{+} - \left(  \alpha_{0}^{-} + \alpha_{1}^{+}\right) &&= 0 -\left( \frac{1}{2} + \frac{1}{2}    \right)  &&= -1\\
d &= \alpha_{\infty}^{+} - \left(  \alpha_{0}^{-} + \alpha_{1}^{-}\right) &&= 0 -\left( \frac{1}{2} + \frac{3}{2}    \right)  &&= -2\\
d &= \alpha_{\infty}^{-} - \left(  \alpha_{0}^{+} + \alpha_{1}^{+}\right) &&= 1 -\left( \frac{1}{2} + \frac{1}{2}    \right)  &&= 0 \\
d &= \alpha_{\infty}^{-} - \left(  \alpha_{0}^{+} + \alpha_{1}^{-}\right) &&= 1 -\left( \frac{1}{2} + \frac{3}{2}    \right)  &&= -1 \\
d &= \alpha_{\infty}^{-} - \left(  \alpha_{0}^{-} + \alpha_{1}^{+}\right) &&= 1 -\left( \frac{1}{2} + \frac{1}{2}    \right)  &&= 0 \\
d &= \alpha_{\infty}^{-} - \left(  \alpha_{0}^{-} + \alpha_{1}^{-}\right) &&= 1 -\left( \frac{1}{2} + \frac{3}{2}    \right)  &&= -1 \\
\end{alignat*} 
There is only one possible $d=0$ values to use. 
Candidate $\omega$ are now found using
\begin{align*} 
\omega&=\sum_{c}\left(  (\pm) \left[  \sqrt{r}\right]_{c}+\frac{\alpha_{c}^{\pm}}{x-c}\right)  +  (\pm)  \left[\sqrt{r}\right]_{\infty}
\end{align*}
Which gives
\begin{align*} 
\omega_1&= \left(  (+) \left[  \sqrt{r}\right]_{0}+\frac{\alpha_{0}^{+}}{x}\right)  
         + \left(  (+) \left[  \sqrt{r}\right]_{1}+\frac{\alpha_{1}^{+}}{x-1}\right)  
         +  (+)  \left[\sqrt{r}\right]_{\infty}\\
\omega_2 &= \left(  (+) \left[  \sqrt{r}\right]_{0}+\frac{\alpha_{0}^{+}}{x}\right)  
         + \left(  (-) \left[  \sqrt{r}\right]_{1}+\frac{\alpha_{1}^{-}}{x-1}\right)  
         +  (+)  \left[\sqrt{r}\right]_{\infty}\\
\omega_3 &= \left(  (-) \left[  \sqrt{r}\right]_{0}+\frac{\alpha_{0}^{-}}{x}\right)  
         + \left(  (+) \left[  \sqrt{r}\right]_{1}+\frac{\alpha_{1}^{+}}{x-1}\right)  
         +  (+)  \left[\sqrt{r}\right]_{\infty}\\
\omega_4 &= \left(  (-) \left[  \sqrt{r}\right]_{0}+\frac{\alpha_{0}^{-}}{x}\right)  
         + \left(  (-) \left[  \sqrt{r}\right]_{1}+\frac{\alpha_{1}^{-}}{x-1}\right)  
         +  (+)  \left[\sqrt{r}\right]_{\infty}\\
\omega_5&= \left(  (+) \left[  \sqrt{r}\right]_{0}+\frac{\alpha_{0}^{+}}{x}\right)  
         + \left(  (+) \left[  \sqrt{r}\right]_{1}+\frac{\alpha_{1}^{+}}{x-1}\right)  
         +  (-)  \left[\sqrt{r}\right]_{\infty}\\
\omega_6 &= \left(  (+) \left[  \sqrt{r}\right]_{0}+\frac{\alpha_{0}^{+}}{x}\right)  
         + \left(  (-) \left[  \sqrt{r}\right]_{1}+\frac{\alpha_{1}^{-}}{x-1}\right)  
         +  (-)  \left[\sqrt{r}\right]_{\infty}\\
\omega_7 &= \left(  (-) \left[  \sqrt{r}\right]_{0}+\frac{\alpha_{0}^{-}}{x}\right)  
         + \left(  (+) \left[  \sqrt{r}\right]_{1}+\frac{\alpha_{1}^{+}}{x-1}\right)  
         +  (-)  \left[\sqrt{r}\right]_{\infty}\\
\omega_8 &= \left(  (-) \left[  \sqrt{r}\right]_{0}+\frac{\alpha_{0}^{-}}{x}\right)  
         + \left(  (-) \left[  \sqrt{r}\right]_{1}+\frac{\alpha_{1}^{-}}{x-1}\right)  
         +  (-)  \left[\sqrt{r}\right]_{\infty}
\end{align*}
Substituting values found in step 1 into the above gives
\begin{alignat*}{3} 
\omega_1&= \left(  \frac{\frac{1}{2}}{x}\right)  + \left(  (+)  \frac{2}{ (x-1)^2}  + \frac{\frac{1}{2}  }{x-1}\right)  &&= \frac{2 x^{2}+x +1}{2 x \left(x -1\right)^{2}}\\
\omega_2 &= \left( \frac{\frac{1}{2}}{x} \right)  + \left(  (-)  \frac{2}{ (x-1)^2} +\frac{\frac{3}{2} }{x-1}\right)  &&=\frac{4 x^{2}-9 x +1}{2 x \left(x -1\right)^{2}}\\
\omega_3 &= \left( \frac{\frac{1}{2}}{x}\right)  + \left(  (+)  \frac{2}{ (x-1)^2} +\frac{\frac{1}{2}}{x-1}\right) &&=\frac{2 x^{2}+x +1}{2 x \left(x -1\right)^{2}} \\
\omega_4 &= \left( \frac{\frac{1}{2}}{x}\right)  + \left(  (-)  \frac{2}{ (x-1)^2} +\frac{\frac{3}{2}}{x-1}\right) &&= \frac{4 x^{2}-9 x +1}{2 x \left(x -1\right)^{2}}\\
\omega_5&= \left(  \frac{\frac{1}{2}}{x} \right)  + \left(  (+) \frac{2}{ (x-1)^2} +\frac{\frac{1}{2}}{x-1}\right)  &&=\frac{2 x^{2}+x +1}{2 x \left(x -1\right)^{2}}\\
\omega_6 &= \left(  \frac{\frac{1}{2}}{x} \right)  + \left(  (-)  \frac{2}{ (x-1)^2} +\frac{\frac{3}{2}}{x-1}\right) &&= \frac{4 x^{2}-9 x +1}{2 x \left(x -1\right)^{2}}\\
\omega_7 &= \left( \frac{\frac{1}{2}}{x}\right)  + \left(  (+) \frac{2}{ (x-1)^2} +\frac{\frac{1}{2}}{x-1}\right)  &&= \frac{2 x^{2}+x +1}{2 x \left(x -1\right)^{2}}\\
\omega_8 &= \left( \frac{\frac{1}{2}}{x}\right)  + \left(  (-) \frac{2}{ (x-1)^2} +\frac{\frac{3}{2}}{x-1}\right)  &&=\frac{4 x^{2}-9 x +1}{2 x \left(x -1\right)^{2}}
\end{alignat*}
Which shows there are only two different $\omega$ to try, these are $\omega_1,\omega_2$.
This complete step 2.  For each trial, step 3 is now invoked. 
Starting with $d=0$ and $\omega=\omega_1=\frac{2 x^{2}+x +1}{2 x \left(x -1\right)^{2}}$. Since the degree $d=0$ then 
$p(x)=1$. This polynomial needs to satisfy
\begin{align*} 
 p''+2 \omega p' + (\omega'+\omega^2-r) p &= 0  \tag{6}\\
(\omega'+\omega^2-r) p &= 0\\
\frac{d}{dx}\left( \frac{2 x^{2}+x +1}{2 x \left(x -1\right)^{2}} \right)
  + \left(\frac{2 x^{2}+x +1}{2 x \left(x -1\right)^{2}}\right)^2- \frac{7 x^{2}+10 x -1}{4 x^{2} \left(x -1\right)^{4}} &=0\\
  0 &= 0
\end{align*}
Because the equation is satisfied, the polynomial $p(x)=1$ can be used.
The solution to $z''=rz$ is now found from
\begin{align*}
z&=p e^{\int \omega\, dx}\\
  &=e^{\int \frac{2 x^{2}+x +1}{2 x \left(x -1\right)^{2}} \, dx}\\
  &=e^{\frac{\ln \left(x -1\right)}{2}-\frac{2}{x -1}+\frac{\ln \left(x \right)}{2}}\\
  &=\sqrt{x -1}\, \sqrt{x}\, {\mathrm e}^{-\frac{2}{x -1}}
\end{align*}
Given this solution for $z(x)$, the first basis solution of the original ode in $y$ is found
using the inverse of the original transformation used to generate the $z$ ode which is
$z = y e^{\frac{1}{2}\int a\,dx}$, therefore 
\begin{align*}
y_1 &= z e^{-\frac{1}{2}\int a\,dx}\\
    &= \sqrt{x -1} \sqrt{x} {\mathrm e}^{-\frac{2}{x -1}} {\mathrm e}^{-\frac{1}{2}\int - \frac{x (3+x) }{x^2 \left(x^2-2 x +1\right)}  \,dx}
\end{align*}
Which simplifies to 
\begin{align*} 
y_1 &= \frac{x^2}{x-1} {\mathrm e}^{-\frac{4}{x -1}}
\end{align*}
The second solution $y_2$ to the original ode is found using 
reduction of order as was done in the first example.
\subsubsection{Example 3}
This ode is a standard second order representing the 
oscillating harmonics ode with constant coefficients and does not 
require Kovacic algorithm to solve it as it can be readily solved 
using standard method by finding the roots of the characteristic equation. It is included here in order to illustrate
the Kovacic algorithm.
\begin{align*}
y''+y'+y&=0\\
A y''+ B y' + C y&=0
\end{align*}
Converting it to $z''=r z$ as shown before gives
\begin{align*}
z''&=\frac{s}{t} z\\
   &=\frac{-3}{4} z
\end{align*}
Hence $r=\frac{-3}{4}$.  There are no poles therefore $\Gamma=\{\}$, and 
$\mathcal{O}(\infty)=\deg(t)-\deg(s)=0$. Table~\ref{tab:first} shows that 
the necessary conditions for case one are only satisfied. 
Since the set $\Gamma$ is empty, then only the quantities related to $\mathcal{O}(\infty)$ need
to be calculated. The order of $r$ at $\infty$ is $O_r(\infty) = 0$ therefore $v=0$.
$r$ has no $x$ in it, hence the Laurent series of $\sqrt r$ at $\infty$ is itself
\begin{align*}
  \sqrt r =\frac{i \sqrt{3}}{2}
\end{align*}
Therefore
\begin{align*}
   a = \frac{i \sqrt{3}}{2}
\end{align*}
And since $r$ is constant then $b=0$. Hence
\begin{alignat*}{3}
       [\sqrt r]_\infty      &=  \frac{i \sqrt{3}}{2}\\ 
       \alpha_{\infty}^{+}   &=  \frac{1}{2} \left( \frac{b}{a} - v \right) &&= \frac{1}{2} \left( \frac{0}{\frac{i \sqrt{3}}{2}} - 0 \right) &&= 0\\ 
      \alpha_{\infty}^{-}   &=  \frac{1}{2} \left( -\frac{b}{a} - v \right) &&= \frac{1}{2} \left( -\frac{0}{\frac{i \sqrt{3}}{2}} - 0 \right) &&= 0
\end{alignat*}
This completes step 1 for case one. Step 2 searches for non-negative integer $d$ using
\begin{align*} 
d&=\alpha_{\infty}^{\pm}-\sum_{c\in \Gamma}\alpha_{c}^{\pm}
\end{align*} 
Since there are no poles then
\begin{align*}
d &= \alpha_\infty^{-} \\ 
  &= 0
\end{align*}
Since $d$ is non-negative integer integer it can be used to find $\omega$ using
\begin{align*}
\omega &= \sum_{c \in \Gamma} \left( s(c) [\sqrt r]_c + \frac{\alpha_c^{s(c)}}{x-c}  \right) + s(\infty) [\sqrt r]_\infty 
\end{align*}
The above reduces to
\begin{align*}
\omega &= (-) [\sqrt r]_\infty \\ 
       &= 0 + (-) \left( \frac{i \sqrt{3}}{2} \right) \\ 
       &= -\frac{i \sqrt{3}}{2}
\end{align*}
Now that $\omega$ is determined, the next step is find a corresponding minimal polynomial $p(x)$ of degree $d=0$ to solve the ode.
The polynomial $p(x)$ needs to satisfy the equation
\begin{align*}
p'' + 2 \omega p' + \left( \omega ' +\omega^2 -r\right) p &= 0\tag{1}
\end{align*}
Since $d=0$ then let $p(x) = 1$. Substituting this in the above gives
\begin{align*}
\left(0\right) + 2 \left(-\frac{i \sqrt{3}}{2}\right) \left(0\right) + \left( \left(0\right) + \left(-\frac{i \sqrt{3}}{2}\right)^2 - \left(-{\frac{3}{4}}\right) \right) &= 0\\ 
0 &= 0
\end{align*}
The equation is satisfied. Therefore the first solution to the ode $z'' = r z$ is 
\begin{align*}
z(x) &= p e^{ \int \omega \,dx} \\ 
            &= {\mathrm e}^{\int -\frac{i \sqrt{3}}{2}d x}\\ 
            &= {\mathrm e}^{-\frac{i \sqrt{3}\, x}{2}}
\end{align*} 
The first solution to the original ode in $y$  is now found from (using $A=1,B=1$)
\begin{align*}
y_1 &= z e^{ \int -\frac{1}{2} \frac{B}{A} \,dx}\\
    &= z e^{ -\int \frac{1}{2} \,dx}\\
    &= z e^{-\frac{x}{2}}\\
    &= {\mathrm e}^{-\frac{i \sqrt{3}\, x}{2}} \left({\mathrm e}^{-\frac{x}{2}}\right)\\
    &={\mathrm e}^{-\frac{x \left(1+\mathrm{I} \sqrt{3}\right)}{2}}
\end{align*}

The second solution $y_2$ to the original ode is found using reduction of order
\begin{align*}
y_2 &= y_1 \int \frac{  e^{\int -\frac{B}{A} \,dx}}{y_1^2} \,dx \\
    &= y_1 \int \frac{  e^{-\int \,dx}}{\left(y_1\right)^2} \,dx\\
    &= y_1 \int \frac{ e^{-x}}{\left(y_1\right)^2} \,dx\\
    &= \left({\mathrm e}^{-\frac{x \left(1+\mathrm{I} \sqrt{3}\right)}{2}}\right) \left(-\frac{i \sqrt{3}\, {\mathrm e}^{i \sqrt{3}\, x}}{3}\right)\\
    &=-\frac{\mathrm{I}}{3} {\mathrm e}^{\frac{x \left(\mathrm{I} \sqrt{3}-1\right)}{2}} \sqrt{3}
\end{align*}
Therefore the general solution is 
\begin{align*}
y &= c_1 y_1 + c_2 y_2\\
  &= c_1 \left({\mathrm e}^{-\frac{x \left(1+i \sqrt{3}\right)}{2}}\right) + c_2 \left(-\frac{i {\mathrm e}^{\frac{x \left(i \sqrt{3}-1\right)}{2}} \sqrt{3}}{3}\right)
\end{align*}
Using Euler's formula the above can be simplified to the standard looking solution
\begin{align*}
y(x) &= {\mathrm e}^{-\frac{x}{2}} \left(  C_1 \sin\left(\frac{\sqrt{3}\, x}{2}\right) + C_2 \cos\left(\frac{\sqrt{3}\, x}{2}\right) \right)
\end{align*}

\subsection{case two}
\subsubsection{Example 1}
Given the ode
\begin{align*}
2 x^2 y''-x y'+(1+x) y&=0
\end{align*}
Converting it $y''+ a y' + b y=0$ gives
\begin{align*}
y''-\frac{1}{2 x} y'+\frac{1+x}{2 x^2} y&=0
\end{align*}
Where $a=-\frac{1}{2 x}, b=\frac{1+x}{2 x^2}$. 
Applying the transformation $z=y e^{\frac{1}{2} \int{a\,dx}}$ gives $z''=r z$ and
$r = \frac{1}{4} a^2 + \frac{1}{2} a' - b$. Therefore
\begin{align*}
r &=\frac{s}{t} \\
  &=\frac{-3-8 x}{16 x^2}
\end{align*}
There is one  pole at $x=0$ of order $2$ and $\mathcal{O}(\infty)=\deg(t)-\deg(s)=2-1=1$. 
Table~\ref{tab:first} shows that necessary conditions for only case two are satisfied. 
Since pole is order $2$, the set $E_0=\{2, 2+ 2\sqrt{1+ 4b}, 2-2 \sqrt{1+ 4b}\}$ 
where $b$ is the coefficient of $\frac{1}{(x)^2}$ in the partial fraction decomposition of $r$ given by
\begin{align*}
r &= -\frac{3}{16 x^{2}}-\frac{1}{2 x}
\end{align*}
Therefore $b=-{\frac{3}{16}}$. Hence
\begin{align*}
E_0 &=\left\{ 2, 2+ 2\sqrt{1- 4 \frac{3}{16}}, 2-2 \sqrt{1- 4 \frac{3}{16}}\right\} \\
    &=\left\{ 2,3, 1 \right\}
\end{align*}
Since $\mathcal{O}(\infty)=1<2$ then $E_{\infty}=\mathcal{O}(\infty)=\{1\}$. This completes step $1$ of case 
two. Step $2$ is used to determine a non-negative integer $d$. Using $e_\infty=1$ gives
\begin{align*} 
d&=\frac{1}{2} \left( e_{\infty} - \sum_{c\in \Gamma} e_c \right) 
\end{align*} 
There is only one pole, so the sum contains only one term. There are 3 possible combinations to try, 
using either $e_0=2$, $e_0=3$ or $e_0=1$.  Therefore
\begin{alignat*}{3}
d&=\frac{1}{2} \left( e_\infty - ( e_0 ) \right)   &&=\frac{1}{2} \left( 1- 2 \right) &&= - \frac{1}{2}\\
  &=\frac{1}{2} \left( e_\infty - ( e_0 ) \right) &&=\frac{1}{2} \left( 1- 3 \right) &&= -1\\
  &=\frac{1}{2} \left( e_\infty - ( e_0 ) \right) &&=\frac{1}{2} \left( 1- 1 \right) &&= 0
\end{alignat*} 
The above shows that only the family $\{e_\infty=1,e_0=1\}$ generated non-negative $d=0$. 
$\theta$ is now found. In the following sum, only $e_c$ retained from the above are used. In this 
example, this  will be $e_0=1$ since it is the member of $E_0$ which generated non-negative integer. If there
were more than one $e_i$ found, then each would be tried at time.
\begin{align*} 
\theta &= \frac{1}{2} \sum_{c\in \Gamma} \frac{e_c}{x-c}\\
       &= \frac{1}{2} \left( \frac{e_0}{x-0} \right)\\
       &=  \frac{1}{2 x}
\end{align*}  
This completes step $2$. Step $3$ finds polynomial $p(x)=a_0+a_1 x+ a_2 x^2 + \dots + x^d$ of degree $d$. 
Since $d=0$ then $p(x)=1$. This polynomial has to satisfy the following
\begin{align*} 
p'''+3 \theta p'' \left(3 \theta^2 + 3 \theta' - 4 r \right) p' + \left(\theta''+ 3 \theta \theta' + \theta^3 - 4 r \theta -2 r' \right)p=0 
\end{align*}  
Substituting $p=1, \theta=\frac{1}{2 x}$ into the above and simplifying gives
\begin{align*} 
0 &= 0
\end{align*}  
Since $p(x)=1$ is verified, then
\begin{align*}  
\phi = \theta + \frac{p'}{p} \\
     = \frac{1}{2 x}
\end{align*} 
Next, $\omega$ solution is found using
\begin{align*} 
\omega^2 - \phi \omega + \left(\frac{1}{2} \phi' +\frac{1}{2} \phi^2 -r\right) &=0 
\end{align*} 
Substituting the values for $\phi$ and $r$ into the above gives
\begin{align*}
w^{2}-\frac{w}{2 x}+\frac{1+8 x}{16 x^{2}} = 0
\end{align*}
Solving for $\omega$ gives two roots, either one can be used. Using 
\begin{align*}
\omega &= \frac{1+2 \sqrt{2}\, \sqrt{-x}}{4 x}
\end{align*}
Therefore the first solution to the ode $z'' = r z$ is 
\begin{align*}
  z(x) &= e^{ \int \omega \,dx} \\ 
              &= {\mathrm e}^{\int \frac{1+2 \sqrt{2}\, \sqrt{-x}}{4 x}d x}\\ 
              &= x^{\frac{1}{4}} {\mathrm e}^{\sqrt{2}\, \sqrt{-x}}
\end{align*} 
The first solution to the original ode in $y$  is now found from
\begin{align*}
y_1 &= z e^{ \int -\frac{1}{2} a \,dx}\\
    &= z e^{ -\int \frac{1}{2} \frac{-x}{2 x^{2}} \,dx}\\
    &= z e^{\frac{\ln \left(x \right)}{4}}\\
    &= \left(x^{\frac{1}{4}} {\mathrm e}^{\sqrt{2}\, \sqrt{-x}}\right) x^{\frac{1}{4}}\\
    &=  \sqrt{x}\, {\mathrm e}^{\sqrt{2}\, \sqrt{-x}} 
\end{align*} 
The second solution $y_2$ to the original ode can be found using reduction of order.
\subsubsection{Example 2}
This is an ode in which the necessary conditions for all three cases are satisfied, but solved
using case two to illustrate the algorithm.
\begin{align*}
(1-x) x^2 y''+(5 x -4) x y'+(6-9 x) y=0
\end{align*}
Converting it $y''+ a y' + b y=0$ gives
\begin{align*}
 y''+\frac{5 x -4}{(1-x) x} y'+\frac{6-9 x}{(1-x) x^2} y=0
\end{align*}
Where $a=\frac{5 x -4}{(1-x) x}, b=\frac{6-9 x}{(1-x) x^2}$. 
Applying the transformation $z=y e^{\frac{1}{2} \int{a\,dx}}$ gives $z''=r z$ where
$r = \frac{1}{4} a^2 + \frac{1}{2} a' - b$. Therefore
\begin{align*}
r &= \frac{s}{t} \\
  &=  \frac{4-x }{4 x (x -1)^2}
\end{align*}
There is a pole at $x=0$ of order $1$ and a pole at $x=1$ of order $2$.  Since there is no odd 
order pole larger than $2$ and the order at $\infty$ is $2$ then the necessary conditions for case one are satisfied.
Since there is a pole of order $2$ then the necessary conditions for case two are also satisfied. Since pole 
order is not larger than $2$ and the order at $\infty$ is $2$ then the necessary conditions 
for case three are also satisfied. Any one of the three cases algorithm could be used to solve this, but
here case two will be used for illustration.

The pole of order $1$ at $x=0$ gives $E_0=\{4\}$ and the pole of order $2$ at $x=1$ gives
$E_1=\{2, 2+ 2\sqrt{1+ 4b}, 2-2 \sqrt{1+ 4b}\}$ where $b$ is the coefficient of $\frac{1}{(x-1)^2}$ in 
the partial fraction decomposition of $r$.  The partial fractions decomposition of $r$ is
\begin{align*}
r = \frac{3}{4 \left(x -1\right)^{2}}-\frac{1}{x -1}+\frac{1}{x}
\end{align*}
The above shows that $b={\frac{3}{4}}$, therefore $E_1 =\{-2, 2, 6\}$.

$\mathcal{O}(\infty)=2$ therefore $E_{\infty}=\{2, 2+ 2\sqrt{1+ 4b}, 2-2 \sqrt{1+ 4b}\}$ where 
$b=\frac{\operatorname{lcoef}(s)}{\operatorname{lcoeff}(t)}$ where $r=\frac{s}{t}$. 
This gives $b=-\frac{1}{4}$, hence
\begin{align*}
E_{\infty}&=\{2, 2+ 2\sqrt{1+ 4b}, 2-2 \sqrt{1+ 4b}\}\\
          &=\left\{2, 2+ 2\sqrt{1- 4\frac{1}{4}}, 2-2 \sqrt{1- 4-\frac{1}{4}}\right\}\\
          &=\{2, 2, 2\}\\
          &=\{2\}
\end{align*}
\begin{center}
\begin{minipage}{\textwidth}
The following table summarizes step $1$ results.
\begin{table}[H]
\centering
\begin{tabular}[c]{|c|c|c|c|c|}\hline
pole $c$ location & pole order & $E_c$  \\ \hline
 $0$ & $1$ & $\{4\}$ \\ \hline
 $1$ & $2$ & $\{-2, 2, 6\}$ \\ \hline
\end{tabular}
\caption{First step, case two. $E_c$ set information}\label{tab:three}
\end{table}
\begin{table}[H]
\centering
\begin{tabular}[c]{|c|c|c|c|}\hline
Order of $r$ at $\infty$ &  $E_\infty $ \\ \hline
$2$ & $\{2\}$ \\ \hline 
\end{tabular}
\caption{First step, case two.  $\mathcal{O}(\infty)$ information}\label{tab:fourth}
\end{table}
\end{minipage}
\end{center}
The above completes step 1 for case two. Step 2 searches for a non-negative integer $d$ using
\begin{align*} 
d&=\frac{1}{2} \left( e_{\infty} - \sum_{c\in \Gamma} e_c \right) 
\end{align*} 
Where in the above $e_c \in E_c$, $e_\infty \in E_\infty$ were found in the first step. The following
are the possible combinations to use
\begin{alignat*}{2}
d&=\frac{1}{2} \left( 2 - ( 4 -2 ) \right) &&= 0 \\
 &=\frac{1}{2} \left(2 - ( 4 +2 )\right) &&=-2 \\
 &=\frac{1}{2} \left(2 - ( 4 +6 )\right) &&=-4 \\
\end{alignat*} 
The above shows that $E_c=\{e_0=4,e_1=-2\}$ are the family of values to use and all other values are discarded. 

The following rational function $\theta$ is determined using
\begin{align*}
\theta &= \frac{1}{2} \sum_{c \in \Gamma} \frac{e_c}{x-c} \\ 
&= \frac{1}{2}  \left(\frac{4}{\left(x-\left(0\right)\right)}+\frac{-2}{\left(x-\left(1\right)\right)}\right) \\ 
&= \frac{2}{x}-\frac{1}{x -1}
\end{align*}
The algorithm now searches for a monic polynomial $p(x)$ of degree $d=0$ such that
\begin{align*}
  p'''+3 \theta p'' + \left(3 \theta^2 + 3 \theta' - 4 r\right)p'     + \left(\theta'' + 3 \theta \theta' + \theta^3 - 4 r \theta - 2 r' \right) p &= 0 
\end{align*}
Since $d=0$, then $p(x)=1$. Substituting the values found in step $2$ in the above equation and simplifying gives
\begin{align*}
0 = 0
\end{align*}
Hence $p(x)=1$ can be used. Let
\begin{align*}
  \phi &= \theta + \frac{p'}{p}\\ 
        &= \frac{2}{x}-\frac{1}{x -1}
\end{align*}
And $\omega$ be the solution of
\begin{align*}
  \omega^2 - \phi \omega + \left( \frac{1}{2} \phi' + \frac{1}{2} \phi^2 - r \right) &= 0
\end{align*}
Substituting the values for $\phi$ and $r$ into the above equation gives
\begin{align*}
w^{2}-\left(\frac{2}{x}-\frac{1}{x -1}\right) w +\frac{\left(x -2\right)^{2}}{4 \left(x -1\right)^{2} x^{2}} &= 0
\end{align*}
Solving for $\omega$ gives two roots, either one can be used. Using 
\begin{align*}
\omega &= \frac{x -2}{2 \left(x -1\right) x}
\end{align*}
Therefore to $z'' = r z$ is 
\begin{align*}
  z  &= e^{ \int \omega \,dx} \\ 
     &= {\mathrm e}^{\int \frac{x -2}{2 \left(x -1\right) x}d x}\\ 
     &= \frac{x}{\sqrt{x -1}}
\end{align*} 
The first solution to the original ode in $y$  is now found from
\begin{align*}
y_1 &= z  e^{- \int \frac{1}{2} a \,dx}\\
    &= z e^{ -\int \frac{1}{2} \frac{5 x^{2}-4 x}{-x^{3}+x^{2}} \,dx}\\
    &= z e^{2 \ln \left(x \right)+\frac{\ln \left(x -1\right)}{2}}\\
    &= z \left(x^{2} \sqrt{x -1}\right)\\
    &=  \frac{x}{\sqrt{x -1}} \left(x^{2} \sqrt{x -1}\right)\\
    &=  x^3
\end{align*}
The second solution $y_2$ to the original ode is found using reduction of order.
\subsection{case three}
\subsubsection{Example 1}
This is the same ode used in second example above for case two as the necessary conditions
for case three are also satisfied.
\begin{align*}
(1-x) x^2 y''+(5 x -4) x y'+(6-9 x) y=0
\end{align*}
As shown earlier, this ode is transformed to $z''=rz$ where
\begin{align*}
r &= \frac{s}{t} \\
  &=  \frac{4-x }{4 x (x -1)^2}
\end{align*}
There is a pole at $x=0$ of order $1$ and a pole at $x=1$ of order $2$.  For
the pole of order $1$ at $x=0$, $E_0=\{12\}$.  For the pole of order $2$ at $x=1$ 
\begin{alignat*}{2}
E_1 &=\left\{ 6 + \frac{12 k}{n} \sqrt{1+ 4 b}\right\} \qquad \text{for} \quad  k&&=-\frac{n}{2}\cdots \frac{n}{2}   \tag{1} 
\end{alignat*} 
Where $k$ is incremented by $1$ each time,  and $n$ is any of $\{4,6,12\}$ and $b$ is the 
coefficient of $\frac{1}{(x-1)^2}$ in the partial fraction decomposition of $r$ which is
\begin{align*}
r = \frac{3}{4 \left(x -1\right)^{2}}-\frac{1}{x -1}+\frac{1}{x}
\end{align*}
The above shows that $b=\frac{3}{4}$. Starting with $n=4$ (if this $n$ produces no solution then $n=6,12$ will be tried also). 
Equation (1) now becomes
\begin{alignat*}{2}
E_1 &=\left\{ 6 + \frac{12 k}{4} \sqrt{1+ 4 \left(\frac{3}{4}\right) } \right\} \qquad \text{for} \quad   k&&=-2\cdots 2 
\end{alignat*}
Which simplifies to
\begin{alignat*}{2}
E_1  &=\left\{ 6 + 6 k  \right\}  \qquad \text{for} \quad   k&&=-2\cdots 2 \\
     &=\left\{ -6, 0, 6, 12, 18 \right\}
\end{alignat*}
$E_\infty$ is found using (1) but with different $b$. In this case $b$ is 
given by $b=\frac{\operatorname{lcoef}(s)}{\operatorname{lcoeff}(t)}$ where $r=\frac{s}{t}$. $\operatorname{lcoef}(s)$ is
the leading coefficient of $s$ and $\operatorname{lcoef}(t)$ is the leading coefficient of $t$. Since
$r=\frac{-x+4 }{4 x^3-8x^2+4 x}$ then  $b = -\frac{1}{4}$. Equation (1) becomes
\begin{alignat*}{2}
E_\infty &=\left\{ 6 + \frac{12 k}{4} \sqrt{1- 4 \left(\frac{1}{4}\right) } \right\} \qquad \text{for} \quad   k&&=-2\cdots 2 
\end{alignat*}
This simplifies to
\begin{alignat*}{2}
E_\infty  &=\left\{6 \right\} 
\end{alignat*}

\begin{center}
\begin{minipage}{\textwidth}
The following table summarizes step $1$ results using $n=4$.
\begin{table}[H]
\centering
\begin{tabular}[c]{|c|c|c|c|c|}\hline
pole $c$ location & pole order & $E_c$  \\ \hline
 $0$ & $1$ & $\{12\}$ \\ \hline
 $1$ & $2$ & $\{-6, 0,6,12,18\}$ \\ \hline
\end{tabular}
\caption{First step, case three using $n=4$. $E_c$ set information}\label{tab:5}
\end{table}
\begin{table}[H]
\centering
\begin{tabular}[c]{|c|c|c|c|}\hline
Order of $r$ at $\infty$ &  $E_\infty $ \\ \hline
$2$ & $\{6\}$ \\ \hline 
\end{tabular}
\caption{First step, case three using $n=4$.  $\mathcal{O}(\infty)$ information}\label{tab:6}
\end{table}
\end{minipage}
\end{center}

The next step is to determine a non negative integer $d$ using
\begin{align*}
d &= \frac{n}{12} \left( e_\infty - \sum_{c \in \Gamma} e_c \right)
\end{align*}
Where in the above $e_c$ is a distinct element from each corresponding $E_c$. This means all possible 
tuples $\{e_{c_1},e_{c_2},\dots,e_{c_n}\}$ are tried in the sum above, where $e_{c_i}$ is one element of each $E_c$ found earlier. 

This results in the following values for $d$ using $n=4$ and $e_\infty=6$.
\begin{alignat*}{2}
d &= \frac{1}{3} \left( 6 - (12 - 6) \right) &&= 0 \\  
  &= \frac{1}{3} \left( 6 - (12 + 0) \right) &&= -2 \\  
  &= \frac{1}{3} \left( 6 - (12 + 6) \right) &&= -4 \\  
  &= \frac{1}{3} \left( 6 - (12 + 12) \right) &&= -6 \\  
  &= \frac{1}{3} \left( 6 - (12 + 18) \right) &&= -8 \\  
\end{alignat*}
Therefore only the first case using $e_0=12,e_1=-6$ generated non-negative integer $d$.
The following rational function is now formed
\begin{align*}
\theta &= \frac{n}{12} \sum_{c \in \Gamma} \frac{e_c}{x-c} \\ 
&= \frac{4}{12}  \left(\frac{12}{\left(x-\left(0\right)\right)}+\frac{-6}{\left(x-\left(1\right)\right)}\right) \\ 
  &= \frac{2 x -4}{\left(x -1\right) x}
\end{align*}
And
\begin{align*}
S &= \prod_{c \in \Gamma} (x-c) \\
  &= (x-0) (x-1) \\
  &= x(x-1)
\end{align*}
This completes the step 2 of the algorithm. 

Since the degree $d=0$, then $p(x)=1$. Now $P_i(x)$ polynomials are generated using
\begin{align*}
P_n &= - p(x) \\ 
P_{i-1} &= - S p'_{i} + ((n-i)S' -S\theta) P_i - (n-1)(i+1) S^2 r P_{i+1} \qquad i=n,n-1,\dots , 0  
\end{align*}
These generate the following set
\begin{align*}
P_{4} &= -1\\ 
P_{3} &= 2 x -4\\ 
P_{2} &= -3 \left(x -2\right)^{2}\\ 
P_{1} &= 3 \left(x -2\right)^{3}\\ 
P_{0} &= -\frac{3 \left(x -2\right)^{4}}{2}\\ 
P_{-1} &= 0
\end{align*}
There is nothing to solve for from the last equation $P_{-1} = 0$ as $p(x)=1$ is already known
because the degree $d$ was zero and hence there are no coefficients $a_i$ to solve for.

Next $\omega$ is determined as the solution to the following equation using $n=4$.
\begin{align*}
 \sum_{i=0}^{n} S^i \frac{P_i}{(n-i)!} \omega^i &= 0 \\
 \frac{P_0}{4!} + \frac{S P_1}{3!} \omega +\frac{S^2 P_2}{2!} \omega^2  +\frac{S^3 P_3}{1!} \omega^3 +  +\frac{S^4 P_4}{0!} \omega^4 &=0\\
 -\frac{1}{16}\left(2 \omega  \,x^{2}-2 x \omega -x +2\right)^{4} &=0
\end{align*}
Solving the above and using any one of the roots results in
\begin{align*}
\omega=\frac{1}{2 x \left(x -1\right)}\left(x -2\right)
\end{align*}
The above $\omega$ is used to find a solution to $z''=r z$ from
\begin{align*}
z &= e^{ \int \omega \,dx}\\
  &= {\mathrm e}^{\int \frac{x -2}{2 x \left(x -1\right)}d x}\\
  &= \frac{x}{\sqrt{x -1}}
\end{align*}
Therefore one solution to the original ode in $y$  is
\begin{align*}
y &= z e^{ \int -\frac{1}{2} a \,dx}\\
  &= z  e^{ -\int \frac{1}{2} \frac{5 x^{2}-4 x}{-x^{3}+x^{2}} \,dx}\\
  &= z e^{2 \ln \left(x \right)+\frac{\ln \left(x -1\right)}{2}}\\
  &= \left(\frac{x}{\sqrt{x -1}}\right)  \left(x^{2} \sqrt{x -1}\right)
\end{align*}
The second solution to the original ode is found using reduction of order.
This completes the solution using case $3$ for degree $n=4$ of $\omega$.

\subsubsection{Example 2}
The ode is 
\begin{align*}
x^2 (1+x) y''+x (2 x +1) y''-(4+6 x) y&=0
\end{align*}
Converting it $y''+ a y' + b y=0$ gives
\begin{align*}
 y''+\frac{2 x +1}{x (1+x)} y'-\frac{4+6 x}{x^2 (1+x)} y=0
\end{align*}
Where $a=\frac{2 x +1}{x (1+x)}, b=-\frac{4+6 x}{x^2 (1+x)}$. 
Applying the transformation $z=y e^{\frac{1}{2} \int{a\,dx}}$ results in $z''=r z$ where
$r = \frac{1}{4} a^2 + \frac{1}{2} a' - b$ where
\begin{align*}
r &= \frac{s}{t} \\
  &=  \frac{24 x^2+40 x +15}{4 \left( x(x+1) \right)^2}
\end{align*}
There is a pole at $x=0$ of order $2$ and a pole at $x=-1$ of order $2$. 
Since there is no odd order pole larger than $2$ and the order at $\infty$ is $2$ then 
the necessary conditions for case one are satisfied. Since there is a pole of order $2$ then necessary 
conditions for case two are also satisfied. Since pole order is not larger than $2$ and the order 
at $\infty$ is $2$ then the necessary conditions for case three are also satisfied.
This is now solved as case three for illustration.

Starting with $n=4$, and for the pole of order $2$ at $x=-1$ 
\begin{alignat*}{2}
E_{-1} &=\left\{ 6 + \frac{12 k}{n} \sqrt{1+ 4 b}\right\} \qquad \text{for} \quad  k&&=-\frac{n}{2}\cdots \frac{n}{2} 
\end{alignat*} 
This simplifies to 
\begin{alignat*}{2}
E_{-1}       &=\left\{ 6 + 3 k \sqrt{1+ 4 b}\right\} \qquad \text{for} \quad  k&&=-2\cdots 2                  \tag{1} 
\end{alignat*} 
$b$ is the coefficient of $\frac{1}{ \left(1+x \right)^{2}}$ in the partial fractions decomposition of $r$ given by
\begin{align*}
r = -\frac{1}{4 \left(1+x \right)^{2}}+\frac{15}{4 x^{2}}-\frac{5}{2 \left(1+x \right)}+\frac{5}{2 x}
\end{align*}
The above shows that $b=-\frac{1}{4}$. Equation (1) becomes
\begin{alignat*}{2}
E_{-1}  &=\left\{ 6 + 3 k \sqrt{1- 4 \left(\frac{1}{4}\right) }\right\} \qquad \text{for} \quad  k&=-2\cdots 2\\  
        &= \{6\}
\end{alignat*}

For the pole at $x=0$ of order $2$, $b$ is the coefficient of $\frac{1}{ x^{2}}$ in the above 
partial fractions decomposition of $r$. This shows that $b={\frac{15}{4}}$. Hence
\begin{alignat*}{2}
E_{0}  &=\left\{ 6 + 3 k \sqrt{1+ 4 \left( \frac{15}{4}\right)  }\right\} \qquad \text{for} \quad  k&=-2\cdots 2 \\
       &=  \{-18, -6, 6, 18, 30\}
\end{alignat*}
$E_\infty$ is found using equation (1) but with different $b$. In this case $b$ is
given by $b=\frac{\operatorname{lcoef}(s)}{\text{lcoeff}(t)}$ where $r=\frac{s}{t}$. $\operatorname{lcoef}(s)$ is
the leading coefficient of $s$ and $\operatorname{lcoef}(t)$ is the leading coefficient of $t$. Since
$r=\frac{24 x^2+40 x +15}{4 x^4+8x^3+4 x^2}$  then  $b =6$. Equation (1) becomes
\begin{alignat*}{2}
E_\infty &=\left\{ 6 + 3 k \sqrt{1+ 4 \left(6\right) } \right\} \qquad \text{for} \quad   k&&=-2\cdots 2 \\
         &=  \{-24, -9, 6, 21, 36\}
\end{alignat*}

\begin{center}
\begin{minipage}{\textwidth}
The following table summarizes step $1$ results using $n=4$.
\begin{table}[H]
\centering
\begin{tabular}[c]{|c|c|c|c|c|}\hline
pole $c$ location & pole order & $E_c$  \\ \hline
 $-1$ & $2$ & $\{6\}$ \\ \hline
 $0$ & $2$ & $  \{-18, -6, 6, 18, 30\}$ \\ \hline
\end{tabular}
\caption{First step, case three using $n=4$. $E_c$ set information}\label{tab:7}
\end{table}
\begin{table}[H]
\centering
\begin{tabular}[c]{|c|c|c|c|}\hline
Order of $r$ at $\infty$ &  $E_\infty $ \\ \hline
$2$ & $ \{-24, -9, 6, 21, 36\}$ \\ \hline 
\end{tabular}
\caption{First step, case three using $n=4$.  $\mathcal{O}(\infty)$ information}\label{tab:8}
\end{table}
\end{minipage}
\end{center}

The next step is to determine a non negative integer $d$ using
\begin{align*}
d &= \frac{n}{12} \left( e_\infty - \sum_{c \in \Gamma} e_c \right)
\end{align*}
Where in the above $e_c$ is a distinct element from each corresponding $E_c$. This means all possible 
tuples $\{e_{c_1},e_{c_2},\dots,e_{c_n}\}$ are tried in the sum above, where $e_{c_i}$ is one element of each $E_c$ found earlier. 

\begin{center}
\begin{minipage}{\textwidth}
This results in the following values for $d$ using $n=4$. 
\footnotesize 
\begin{alignat*}{2}
d &= \frac{1}{3} \left( -24 - (6 -18) \right) &= -4\\
  &= \frac{1}{3} \left( -24 - (6 +6) \right) &=-12\\
  &= \frac{1}{3} \left( -24 - (6 -6) \right) &= -8\\
  &=\frac{1}{3} \left( -24 - (6 +18) \right) &= -16\\
  &= \frac{1}{3} \left( -24 - (6 +30) \right) &= -20\\
  &= \frac{1}{3} \left( -9 - (6 -18) \right) &=1\\
  &= \frac{1}{3} \left( -9 - (6 +6) \right) &=-7\\
  &= \frac{1}{3} \left( -9 - (6 -6) \right) &= -3\\
  &= \frac{1}{3} \left( -9 - (6 +18) \right) &= -11\\
  &= \frac{1}{3} \left( -9 - (6 +30) \right) &= -15\\
  &= \frac{1}{3} \left( 6 - (6 -18) \right) &= 6\\
  &= \frac{1}{3} \left( 6 - (6 +6) \right) &= -2\\
  &= \frac{1}{3} \left( 6 - (6 -6) \right) &=2 \\
  &= \frac{1}{3}\left( 6 - (6 +18) \right) &= -6\\
  &= \frac{1}{3} \left( 6 - (6 +30) \right) &=-10 \\
  &= \frac{1}{3} \left( 21 - (6 -18) \right) &= 11\\
  &= \frac{1}{3} \left( 21 - (6 +6) \right) &= 3\\
  &= \frac{1}{3} \left( 21 - (6 -6) \right) &= 7\\
  &= \frac{1}{3} \left( 21 - (6 +18) \right) &= -1\\
  &= \frac{1}{3} \left( 21 - (6 +30) \right) &= -5\\
  &= \frac{1}{3} \left( 36 - (6 -18) \right) &= 16\\
  &= \frac{1}{3} \left( 36 - (6 +6) \right) &=8\\
  &= \frac{1}{3} \left( 36 - (6 -6) \right) &=12\\
  &=\frac{1}{3} \left( 36 - (6 +18) \right) &= 4\\
  &= \frac{1}{3} \left( 36 - (6 +30) \right) &= 0
\end{alignat*}
\normalsize
\end{minipage}
\end{center}
From the above,  the following families all produce non-negative $d$
\begin{center}
\begin{tabular}{llll}
$e_\infty=-9$ & $e_{-1}=6$ & $e_{0}=-18$ & $d=1$\\
$e_\infty=6$ & $e_{-1}=6$ & $e_{0}=-18$ & $d=6$\\
$e_\infty=6$ & $e_{-1}=6$ & $e_{0}=-6$ & $d=2$\\
$e_\infty=21$ & $e_{-1}=6$ & $e_{0}=-18$ & $d=11$\\
$e_\infty=21$ & $e_{-1}=6$ & $e_{0}=6$ & $d=3$\\
$e_\infty=21$ & $e_{-1}=6$ & $e_{0}=-6$ & $d=7$\\
$e_\infty=36$ & $e_{-1}=6$ & $e_{0}=-18$ & $d=16$\\
$e_\infty=36$ & $e_{-1}=6$ & $e_{0}=6$ & $d=8$\\
$e_\infty=36$ & $e_{-1}=6$ & $e_{0}=-6$ & $d=12$\\
$e_\infty=36$ & $e_{-1}=6$ & $e_{0}=18$ & $d=4$\\
$e_\infty=36$ & $e_{-1}=6$ & $e_{0}=30$ & $d=0$
\end{tabular}
\end{center}

Starting with the smallest $d=0$ as that is the least computationally expensive with its
corresponding $e_{-1}=6,e_0=30,e_\infty=36$, the following rational function is now formed
\begin{align*}
\theta &= \frac{n}{12} \sum_{c \in \Gamma} \frac{e_c}{x-c} \\ 
  &= \frac{4}{12}  \left(\frac{6}{x+1}+\frac{30}{x}\right) \\
  &= \frac{12 x +10}{x \left(1+x \right)}
\end{align*}
And
\begin{align*} 
S &= \prod_{c\in \Gamma} (x-c) \\ 
  &=  (x+1)x
\end{align*}
This completes the step 2 of the algorithm. 

Polynomial $p(x)$ is now determined. Since the degree of the polynomial is $d=0$, then
\begin{align*}
p(x) = 1
\end{align*}
The $P_i(x)$ polynomials are generated using
\begin{align*}
P_n &= - p(x) \\ 
P_{i-1} &= - S p'_{i} + ((n-i)S' -S\theta) P_i - (n-1)(i+1) S^2 r P_{i+1} \qquad i=n,n-1,\dots , 0  
\end{align*}
The above results in the following set
\begin{align*}
P_{4} &= -1\\
P_{3} &= 12 x +10\\
P_{2} &= -3 \left(6 x +5\right)^{2}\\
P_{1} &= 3 \left(6 x +5\right)^{3}\\
P_{0} &= -\frac{3 \left(6 x +5\right)^{4}}{2}\\
P_{-1} &= 0
\end{align*}
There is coefficient for $p(x)$ to solve for from the last equation $P_{-1} = 0$ as $p(x)=1$ is already known
because the degree $d$ is zero. $\omega$ is now determined as the solution to the following equation, using $n=4$
\begin{align*}
 \sum_{i=0}^{n} S^i \frac{P_i}{(n-i)!} \omega^i &= 0 \\
 \frac{P_0}{4!} + \frac{S P_1}{3!} \omega +\frac{S^2 P_2}{2!} \omega^2  +\frac{S^3 P_3}{1!} \omega^3 +  +\frac{S^4 P_4}{0!} \omega^4 &=0\\
  -\frac{1}{16} \left(2 \omega  \,x^{2}+2 x \omega -6 x -5\right)^{4} &=0 
\end{align*}
Solving the above and using any one of the roots gives
\begin{align*}
\omega=\frac{x-2}{2 x \left(x -1\right)}
\end{align*}
This $\omega$ is used to find a solution to $z''=r z$ from
\begin{align*}
z &= e^{ \int \omega \,dx}\\
  &= {\mathrm e}^{\int \frac{6 x +5}{2 x \left(1+x \right)}d x}\\
  &= x^{\frac{5}{2}} \sqrt{1+x}
\end{align*}
The first solution to the original ode in $y$  is found from
\begin{align*}
y &= z e^{ \int -\frac{1}{2} a \,dx}\\ 
  &= z e^{ -\int \frac{1}{2} \frac{2 x^{2}+x}{x^{2} \left(1+x \right)} \,dx}\\
 &= z  e^{-\frac{\ln \left(x \left(1+x \right)\right)}{2}}\\
 &=\left( x^{\frac{5}{2}} \sqrt{1+x}\right) \left(\frac{1}{\sqrt{x \left(1+x \right)}}\right)
\end{align*}
Which simplifies to
\begin{align*}
y &= x^{2}
\end{align*}
The second solution to the original ode is found using reduction of order.
\clearpage
\section{Conclusion}
Detailed description of the Kovacic algorithm with worked out examples
were given. All three cases of the Kovacic algorithm were implemented using 
object oriented design in Maple. The software was then used to analyze 
over $3000$ differential equations. The results showed that case one and two 
combined provided coverage for $99.9$\% of the ode's  with $97.36$\% of the ode's 
solved using case one algorithm and $2.54$\% solved using case $2$ algorithm 
with only $0.1$\% requiring case $3$. Not a single ode was found that 
required the use of case three with $n=6$ or $n=12$. 

One restriction found on the use of the algorithm is that it requires an ode with its
coefficients being numerical and not symbolic. This is because the algorithm  
has to decide in step $2$ if $d$ (the degree of polynomial  $p(x)$) is 
non-negative integer or not in order to continue to step $3$. If some of 
the ode coefficients were symbolic, it will not be able to decide on this
(without additional assumptions provided). Therefore this algorithm works 
best with ode's having its coefficients given with numerical values.

\clearpage
\section{Appendix}
\subsection{Instructions and examples using the Kovacic package}
\lstdefinestyle{maple}{%
  basicstyle=\ttfamily\footnotesize,
  numbers=left,
  stepnumber=1,
  keywordstyle=\color{blue},%
  stringstyle=\color{mylilas},
	breaklines=false,
    columns=fullflexible,
    keepspaces=true,
	backgroundcolor=\color{bg},
	rulecolor=\color{gray},
	language=,
	frame=single,
	frameround=tttt,
	aboveskip=12pt,belowskip=6pt,
  showstringspaces=false,
  morekeywords={and,assuming,break,by,catch,description,do,done,elif,else,
   end,error,export,fi,finally,for,from,global,if,implies,in,intersect,local,minus,mod,module,next,
   not,od,option,options,or,proc,quit,read,return,save,stop,subset,then,to,try,union,until,use,uses,while,xor},%
   morendkeywords={algebraic,anyfunc,anything,atomic,boolean,complex,constant,cx_infinity,
       cx_zero,embedded_axis,embedded_imaginary,embedded_real,equation,even,extended_numeric,extended_rational,
       finite,float,fraction,function,identical,imaginary,indexable,indexed,integer,list,literal,module,
       moduledefinition,name,neg_infinity,negative,negint,negzero,nonnegative,nonnegint,nonposint,nonpositive,
       nonreal,numeric,odd,polynom,pos_infinity,posint,positive,poszero,procedure,protected,radical,range,rational,
       ratpoly,real_infinity,realcons,relation,sequential,set,sfloat,specfunc,string,symbol,tabular,uneval,
       zppoly,ASSERT,Array,ArrayOptions,CopySign,DEBUG,Default0,DefaultOverflow,DefaultUnderflow,ERROR,EqualEntries,
       EqualStructure,FromInert,Im,MPFloat,MorrBrilCull,NameSpace,NextAfter,Normalizer,NumericClass,NumericEvent,
       NumericEventHandler,NumericStatus,Object,OrderedNE,RETURN,Re,Record,SDMPolynom,SFloatExponent,
       SFloatMantissa,Scale10,Scale2,SearchText,TRACE,ToInert,Unordered,UpdateSource,_hackwareToPointer,_jvm,_local,
       _maplet,_savelib,_treeMatch,_unify,_xml,abs,add,addressof,anames,andmap,appendto,
       array,assemble,assign,assigned,attributes,bind,call_external,callback,cat,coeff,coeffs,
       conjugate,convert,crinterp,debugopts,define_external,degree,denom,diff,disassemble,divide,dlclose,
       done,entries,eval,evalb,evalf,evalgf1,evalhf,evalindets,evaln,expand,exports,factorial,frem,frontend,
       gc,genpoly,gmp_isprime,goto,has,hastype,hfarray,icontent,ifelse,igcd,ilog10,ilog2,implies,
       indets,indices,inner,iolib,iquo,irem,is_gmp,isqrt,kernelopts,lcoeff,ldegree,length,lexorder,
       lhs,localGridInterfaceRun,lowerbound,lprint,macro,map,map2,max,maxnorm,member,membertype,
       min,minus,mod,modp,modp1,modp2,mods,mul,mvMultiply,negate,nops,normal,numboccur,numelems,
       numer,op,order,ormap,overload,parse,piecewise,pointto,print,print_preprocess,readlib,
       reduce_opr,remove,rhs,rtable,rtableInfo,rtable_convolution,rtable_eval,rtable_histogram,rtable_indfns,
       rtable_is_zero,rtable_normalize_index,rtable_num_dims,rtable_num_elems,rtable_options,rtable_redim,
       rtable_scale,rtable_scanblock,rtable_size,rtable_sort_indices,rtable_zip,savelib,searchtext,
       select,selectremove,seq,series,setattribute,sign,sort,ssystem,stop,streamcall,subs,subset,
       subsindets,subsop,substring,system,table,taylor,tcoeff,time,timelimit,traperror,trunc,type,
       typematch,unames,unbind,upperbound,userinfo,wbOpen,wbOpenURI,writeto,
       ~Array,~Matrix,~Vector,args,nargs,procname,RootOf,Float,thismodule,thisproc,_options,
      _noptions,_rest,_nrest,_params,_nparams,_passed,_npassed,_nresults,static,Catalan,true,false,
      FAIL,infinity,Pi,gamma,integrate,libname,NULL,Order,printlevel,lasterror,lastexception,Digits,
      constants,undefined,I,UseHardwareFloats,Testzero,Normalizer,NumericEventHandlers,
      Rounding,Catalan,FAIL,Pi,false,gamma,infinity,true,ansi,echo,errorbreak,errorcursor,
      indentamount,labeling,labelwidth,patchlevel,plotdevice,plotoptions,plotoutput,postplot,
      preplot,prettyprint,printbytes,prompt,quiet,screenheight,screenwidth,showassumed,verboseproc,
      version,warnlevel,ASSERT,bytesalloc,bytesused,cputime,dagtag,gcbytesavail,gcbytesreturned,gctimes,
      maxdigits,maximmediate,memusage,printbytes,profile,system,version,wordsize,_Inert,And,Non,Not,Or,
      SERIES,SymbolicInfinity,TEXT,algebraic,algext,algfun,algnum,algnumext,anyfunc,anything,arctrig,
      atomic,boolean,complex,complexcons,constant,cubic,cx_infinity,cx_zero,embedded_axis,
      embedded_imaginary,embedded_real,equation,even,evenfunc,expanded,extended_numeric,
      extended_rational,facint,finite,float,fraction,function,hfloat,identical,imaginary,
      indexable,indexed,infinity,integer,laurent,linear,list,listlist,literal,logical,mathfunc,
      matrix,moduledefinition,monomial,name,neg_infinity,negative,negint,negzero,nonnegative,
      nonnegint,nonposint,nonpositive,nonreal,nothing,numeric,odd,oddfunc,package,point,polynom,
      pos_infinity,posint,positive,poszero,prime,protected,quadratic,quartic,radext,radfun,radfunext,
      radical,radnum,radnumext,range,rational,ratpoly,real_infinity,realcons,relation,scalar,
      sequential,set,sfloat,specfunc,specindex,sqrt,stack,string,symbol,symmfunc,tabular,trig,
      truefalse,truefalseFAIL,undefined,uneval,vector,zppoly,AFactor,AFactors,AiryAi,AiryAiZeros,
      AiryBi,AiryBiZeros,AngerJ,ArrayDims,ArrayElems,ArrayIndFns,ArrayNumDims,Berlekamp,BesselI,
      BesselJ,BesselJZeros,BesselK,BesselY,BesselYZeros,Beta,Cache,ChebyshevT,ChebyshevU,CheckArgs,
      Chi,Ci,Complex,ComplexRange,Content,CoulombF,CylinderD,CylinderU,CylinderV,D,DESol,Describe,
      Det,Diff,Dirac,DistDeg,Divide,Ei,EllipticCE,EllipticCK,EllipticE,EllipticF,EllipticK,
      EllipticPi,Eval,Expand,Explore,ExportVector,Factor,Factors,Fraction,FresnelC,FresnelS,
      Fresnelf,Fresnelg,GAMMA,GF,Gausselim,Gaussjord,Gcd,Gcdex,GegenbauerC,HFloat,HankelH1,HankelH2,
      Heaviside,Hermite,HermiteH,HeunB,HeunBPrime,HeunC,HeunCPrime,HeunD,HeunDPrime,HeunG,
      HeunGPrime,HeunT,HeunTPrime,ImportVector,Indep,Int,Intat,Interp,InverseJacobiAM,InverseJacobiCD,
      InverseJacobiCN,InverseJacobiCS,InverseJacobiDC,InverseJacobiDN,InverseJacobiDS,InverseJacobiNC,
      InverseJacobiND,InverseJacobiNS,InverseJacobiSC,InverseJacobiSD,InverseJacobiSN,Irreduc,
      IsMatrixShape,IsVectorShape,IsWorksheetInterface,JacobiAM,JacobiCD,JacobiCN,JacobiCS,JacobiDC,
      JacobiDN,JacobiDS,JacobiNC,JacobiND,JacobiNS,JacobiP,JacobiSC,JacobiSD,JacobiSN,JacobiTheta1,
      JacobiTheta2,JacobiTheta3,JacobiTheta4,KelvinBei,KelvinBer,KelvinHei,KelvinHer,KelvinKei,KelvinKer,
      KummerM,KummerU,LaguerreL,Lcm,LegendreP,LegendreQ,Limit,LommelS1,LommelS2,MOLS,MathieuA,MathieuB,MathieuC,
      MathieuCE,MathieuCEPrime,MathieuCPrime,MathieuExponent,MathieuFloquet,MathieuFloquetPrime,MathieuS,
      MathieuSE,MathieuSEPrime,MathieuSPrime,Matrix,MatrixOptions,MeijerG,Normal,Nullspace,Power,Powmod,
      Prem,Primfield,Primitive,Product,Psi,Quo,RESol,Randpoly,Ratrecon,RealRange,Rem,Resultant,Roots,Shi,
      Si,Smith,Sqrfree,Ssi,Stirling1,Stirling2,String,StruveH,StruveL,Sum,TopologicalSort,Trace,Vector,
      WARNING,WeberE,WeierstrassP,WeierstrassPPrime,WeierstrassSigma,WeierstrassZeta,WhittakerM,WhittakerW,
      Wrightomega,about,addcoords,additionally,addproperty,algsubs,alias,allvalues,andseq,apply,applyop,
      applyrule,arccos,arccosh,arccot,arccoth,arccsc,arccsch,arcsec,arcsech,arcsin,arcsinh,arctan,
      arctanh,assume,asympt,bernoulli,bernstein,binomial,branches,ceil,charfcn,chrem,coeftayl,collect,
      combine,comparray,compiletable,compoly,content,convergs,copy,cos,cosh,cot,coth,coulditbe,csc,csch,
      dataplot,define,definemore,depends,dilog,dinterp,discont,discrim,dismantle,dsolve,eliminate,ellipsoid,
      erf,erfc,erfi,euler,eulermac,evala,evalapply,evalc,evalr,evalrC,example,exists,exp,extrema,factor,
      factors,fdiscont,fixdiv,floor,fnormal,forall,forget,frac,freeze,fsolve,galois,gcd,gcdex,getassumptions,
      hasassumptions,hasfun,hasoption,help,history,identify,ifactor,ifactors,igcdex,ilcm,ilog,implicitdiff,index,
      info,initialcondition,insertpattern,int,intat,interp,intsolve,invfunc,invztrans,iperfpow,iratrecon,
      iroot,irreduc,is,iscont,isolate,isolve,ispoly,isprime,isqrfree,issqr,ithprime,latex,lcm,leadterm,limit,
      ln,lnGAMMA,log,log10,log2,maptype,match,maximize,minimize,modpol,msolve,mtaylor,nextprime,norm,
      nprintf,odetest,orseq,packages,patmatch,plot,plot3d,plotsetup,poisson,polylog,powmod,prem,prevprime,
      primpart,printf,product,proot,protect,psqrt,quo,radfield,radnormal,rand,randomize,randpoly,
      rationalize,ratrecon,readdata,readstat,realroot,redefine,reduce,related,rem,residue,resultant,
      root,rootbound,roots,round,rsolve,rtable_dims,rtable_elems,scanf,sec,sech,selectfun,shake,
      showtime,signum,simplify,sin,singular,sinh,sinterp,smartplot,smartplot3d,solve,sprem,sprintf,
      sqrfree,sscanf,sturm,sturmseq,subtype,sum,surd,symmdiff,tablelook,tan,tanh,testeq,thaw,
      trigsubs,unapply,unassign,undefine,unprotect,unwindK,usage,value,verify,version,whattype,
      xormap,xorseq,ztrans,interface,readline,with,unwith}%
  otherkeywords={\%,\%\%,\%\%\%,\$define,\$elif,\$else,\$endif,\$file,\$ifdef,\$ifndef,\$include,\$undef},%
  sensitive=true,%
  morestring=[b]",%
  morestring=[b]`,%
  morecomment=[l]\#,%
  morecomment=[s]{(*}{*)}%
  }[keywords,comments,strings]

The Kovacic class is included in the file \verb|KOV.mpl| and the Kovacic testsuite module is 
in the file \verb|kovacic_tester.mpl|. These two files accompany the arXiv version of this paper.

To use these, download these two files to some directory at your computer. For example, on windows, 
assuming the files were downloaded to \verb|c:/my_folder/|, then now start Maple and type
\begin{verbatim}
read "c:/my_folder/KOV.mpl"
read "c:/my_folder/kovacic_tester.mpl"
\end{verbatim}
The above will load the \verb|kovacic_class| and the testsuite module. Once 
the above is successfully completed, then to solve an ode the command is
\begin{lstlisting}[style=maple]
ode  := diff(y(x),x$2)+diff(y(x),x)+y(x)=0;
o    := Object(kovacic_class,ode,y(x)); #create the object
sol  := o:-dsolve();
\end{lstlisting}

The above command will automatically try all the cases that have been
detected one by one until a solution is found. If no solution is found, it
returns \verb|FAIL|. To verify the solution, the command is
\begin{lstlisting}[style=maple]
if sol<>FAIL then
   odetest(sol,ode);
fi;
\end{lstlisting}
Which returns $0$ if the solution is correct.

A note on the type of ode's supported: it is recommenced to use only ode's with 
numeric coefficients and not symbolic coefficients. This is because the 
Kovacic algorithm needs to decide if the degree $d$ of the 
polynomial $p(x)$ is non-negative or not. If some of the coefficients 
are purely symbolic, then it can fail to decide this. An example of 
this is given in the original Kovacic paper as example 2 on page 14, which is to solve
the Bessel ode $y'' =  \left(\frac{4 n^{2}-1}{4 x^{2}}-1\right) y$. This will 
now return \verb|FAIL| since the algorithm can not decide if $d$ is non-negative 
integer without knowing any assumptions or having numerical value for $n$. 
Replacing $n$ by any half odd integer, then it can solve it as follows
\begin{lstlisting}[style=maple]
ode :=diff(diff(y(x),x),x)=((4*n^2-1)/(4*x^2)-1)*y(x);
n   :=-3/2;
o   := Object(kovacic_class,ode,y(x));
sol := o:-dsolve();
\end{lstlisting}

To solve an ode using specific case number, say case $2$, the command is
\begin{lstlisting}[style=maple]
ode   := ...;
o     := Object(kovacic_class,ode,y(x));
sol   := o:-dsolve_case(2);
\end{lstlisting}

If the ode happened to satisfy cases $1$ and $2$ for an example, then 
the above command will only use case $2$ to solve it and will skip case $1$. If 
the command \verb|o:-dsolve()| was used instead, then the ode will be solved 
using case $1$ instead as that is the first one tried. Case $2$ will only be
tried is no solution is found using case $1$.

The object created above, named as ``o'', has additional public methods that 
can be invoked. The following is description of all public methods available.
\begin{itemize}
\item \verb|o:-get_y_ode()| This returns the original ode.
\item \verb|o:-get_z_ode()| This returns the ode solved by Kovacic algorithm which is $z''=r z$.
\item \verb|o:-get_r()| This returns $r$ only.
\item \verb|o:-get_poles()| This returns list of the poles of $r$. It has
the format 

\verb|[ [pole location,pole order],[pole location,pole order], ...]|

If there are no poles, then the empty list \verb|[]| is returned.
\item \verb|o:-get_order_at_infinity()| This returns the order of $r$ at infinity.
\item \verb|o:-get_possible_cases()| This returns list of possible Kovacic cases detected which
can be \verb|[1]|, \verb|[2]|, \verb|[1,2]|, \verb|[1,2,3]|. If no Kovacic cases are
found, then the empty list \verb|[]| is returned.
\item \verb|o:-get_case_used()| This returns the actual case number used if solution to the 
ode was successful. This can be $1,2$ or $3$. If no solution is found after trying all
cases whose conditions were satisfied, then $-1$ is returned.
\item \verb|o:-get_n_case_3()| This return $n$ used when case $3$ was used to solve
the ode. This can be $4,6$ or $12$. If case $3$ was not used, or no solution is
found, then $-1$ is returned.
\end{itemize}

To run the full testsuite of $3000$ ode's that comes with the package, the command is
\begin{lstlisting}[style=maple]
kovacic_tester:-unit_test_main_api();
                   "Test ", 6735, " PASSED "
                   "Test ", 6736, " PASSED "
                   "Test ", 6737, " PASSED "
                             .
                             .
                   "Test ", 7579, " PASSED "
                   "Test ", 7580, " PASSED "
                   "Test ", 7581, " PASSED "
\end{lstlisting}
To run testsuite using specific cases only, the commands are
\begin{lstlisting}[style=maple]
kovacic_tester:-unit_test_case_1();
kovacic_tester:-unit_test_case_2();
kovacic_tester:-unit_test_case_3();
\end{lstlisting}

\clearpage
\subsection{Source code}

\begin{lstlisting}[style=maple]
#--------------------------------------------------------------------
#FILE NAME  :   KOV.mpl
#
# Copyright (C) 2022, Nasser M. Abbasi. All rights reserved
# email: nma@12000.org
#
# Free software to use and modify in anyway as long as the above 
# copyright notice remains attached in the file
#
# Change history
#-----------------
# Oct 27, 2022.     Initial version
#
# Note that the latest version and any updates can alaways be obtained 
# from the author web site at 
# http://12000.org/my_notes/my_paper_on_kovacic/paper.htm
#
# Any problems found in the software please report so I can correct.
# This implementation was done using Maple 2022.2 on windows 10.
#
#--------------------------------------------------------------------

#--------------------------------------------------------------------
# An Object oriented implementation of the Kovacic algortithm 
# using Maple 2022.
#
# based on original Kovacic paper description of the algorithm.
# 
# This file contains two modules. The first is called kovacic_class 
# used to create kovacic object. The second is kovacic_tester module 
# used for unit testing the kovacic_class module
#
# To use this file just do
#
#    read "KOV.mpl"
#    ode := diff(y(x),x$2)=2/x^2*y(x)
#    o   := Object(kovacic_class,ode,y(x));
#    sol := o:-dsolve();
#
# Make sure to set the currentdir() in Maple correctly in order to find 
# where you downloaded the file "KOV.mpl" to.
#--------------------------------------------------------------------

#--------------------------------------------------------------------
# This class is used to solve an ode using Kovacic algorithm.
#
#  Please see documantion section in the paper for additional 
#  information on using this class.
#--------------------------------------------------------------------

kovacic_class :=module()
option object;

#class to hold one entry in the gamma set for case 1
local case_one_gamma_entry := module()
    option object;
    export pole_location := 0;
    export pole_order    := 0;
    export sqrt_r        := 0;
    export alpha_plus    := 0;
    export alpha_minus   := 0; 
    export b:=0;
end module;

#class to hold O_infinity information for case 1
local case_one_O_inf := module()
    option object;
    export sqrt_r_inf      := 0;
    export alpha_plus_inf  := 0;
    export alpha_minus_inf := 0;
    export a := 0;
    export b := 0;
end module;

#class to hold one entry in the gamma set for case 2,3
local case_2_and_3_gamma_entry:=module()
   option object;

   export pole_location := 0;
   export pole_order    := 0;
   export Ec::set       := {};
   export b := 0;
end module;

#--------------------------------------------------------------------
#  PRIVATE variables for the kovacic class
#  only methods inside this class can access these
#--------------------------------------------------------------------

local original_ode; 

#coefficients of original linear ode A y''+ B y' + C y =0    
local A,B,C;

#original ode dependent and independent variables
local y::symbol,x::symbol;

local modified_ode;  #this is the z''=r*z    ode;
local z::symbol; #dependent variable for the r_ode z''(x)=r*z(x)
local r,s,t;     #where r=s/t;

local O_inf; #order of r at infinity. degree(s)-degree(t)

# poles or r. format is [ [pole,order],[pole,order],...]
# if no poles, for example r=x, then list is empty []
# this means pole order zero is not in the list, since no pole.
local poles_list::list := [];  

#contains all possible cases detected. Hence [1] or [1,2] or [1,2,3] 
#or remains empty [] for case 4.
local list_of_possible_cases::list := [];

local case_used_to_solve::integer:=-1; #this will have case used: 1,2 or 3

#this will have n=4,6 or 12 used for case 3 only
local n_used_for_case_3::integer:=-1; 

#--------------------------------------------------------------------
#  CONSTRUCTOR
#
#  the input is  the ode itself as first argument, and the 
#  dependent variable, y(x) for example, as second argument.
#--------------------------------------------------------------------
export ModuleCopy::static:= proc( _self, 
            proto::kovacic_class, 
            ode::`=`,func::function(name) ,$)
    local A,B,C;
    local x,y,r,z::nothing,s,t;

    x,y,A,B,C := parse_and_validate_ode(ode,func);

    _self:-original_ode := ode;

    #force r to be relatively prime ratio of 2 polynomials    
    r := normal( (2*diff(B,x)*A-2*B*diff(A,x)+B^2-4*A*C)/(4*A^2));
    if not type(r,'ratpoly'(anything,x)) then 
        ERROR("r= ", r ," is not polynomial or rational function in ",x); 
    fi; 

    s := numer(r);
    t := denom(r);

    #save all findings into the object private data
    #so it can be used later
    _self:-modified_ode := diff(z(x),x$2) = r*z(x);
    _self:-O_inf := degree(t,x)-degree(s,x); 

    _self:-r  := r;
    _self:-x  := x;
    _self:-y  := y;
    _self:-z  := z;
    _self:-s  := s;
    _self:-t  := t;
    _self:-C  := C; 
    _self:-B  := B; 
    _self:-A  := A; 

    #determine all poles and order
    _self:-generate_poles_and_order_list();

    #determine all possible kovacic cases
    _self:-find_possible_cases();

    return NULL;
end proc;

#--------------------------------------------------------------------
# This module private function is called by constructor to parse and
# validate the ode
#--------------------------------------------------------------------
local parse_and_validate_ode:=proc(ode::`=`,func::function(name),$)
    local x,y,A,B,C,L::list;
    local item,dep_variables_found;

    if nops(func)<>1 then
        ERROR("dependent variable ",func," has more than one argument");
    fi;

    y := op(0,func);
    x := op(1,func);

    #basic verification
    if not has(ode,y) then ERROR("Supplied ",ode," has no ",y); fi;
    if not has(ode,x) then ERROR("Supplied ", ode," has no ",x); fi;
    if not has(ode,func) then ERROR("Supplied ",ode," has no ",func); fi;

    #check it is second order
    if PDEtools:-difforder(ode)<>2 then
        ERROR("Only second order ode's can be used in Kocacic algorithm. ");
    fi;

    #check all dependent variables in ode which is y, match given func y(x)
    try    
        dep_variables_found := PDEtools:-Library:-GetDepVars([y],ode);
    catch:
        ERROR(lastexception);
    end try;

    #o over dep_variables_found and check the 
    #independent variable is same as x i.e. ode can be y'(z)+y(z)=0 but 
    #function is y(x).     
    for item in dep_variables_found do
        if not type(item,function) then
            ERROR("Parse error. Expected ",func," found ",item," in", ode);
        else
            if op(1,item) <> x then
                ERROR("Parse error. Expected ",func," found ",item," in",ode);
            fi;
            if nops(item)<>1 then
                ERROR("Parse error. One argument allowed in ",
                      func," found ", item," in " , ode);
            fi;
        fi;
    od;     

    #now go over all indents in ode and check that y shows as y(x) 
    #and not as just y as the PDEtools:-Library:-GetDepVars([_self:-y],ode) 
    #code above does not detect this. i.e. it does not check  y'(x)+y=0
    if numelems(indets(ode,identical(y))) > 0 then
       ERROR("Parsing error, Can not have ",y," with no argument inside ",ode);
    fi; 

    #check ode is linear in y(x)
    if not has(DEtools:-odeadvisor(ode,func,['linear']),'_linear') then
        ERROR("Only linear ode's can be used in Kocacic algorithm. ");
    fi;

    #extract coefficients of the ode
    L:=DEtools:-convertAlg(ode,func); #this only works on linear ode's

    if L[2]<>0 then ERROR("Not homogeneous ode"); fi;

    #Finished parsing the ode.  A y''+B y' + C y = 0
    C  := L[1,1]; 
    B  := L[1,2]; 
    A  := L[1,3]; 

    if not type(A,'ratpoly'(anything,x)) or  not type(B,'ratpoly'(anything,x)) 
        or not type(C,'ratpoly'(anything,x)) then
        error "ode coefficients are not rational functions of ",x;
    fi;

    return x,y,A,B,C;

end proc;

#--------------------------------------------------------------------
# main API.  Called to solve the ode using Kovacic algorithm.
# returns the solution in the form of y(x)=.... or if no solution
# is found returns FAIL.
# see user guide how to use.
#--------------------------------------------------------------------
export dsolve::static:=proc(_self,$)
    local sol:=FAIL, current_case_number::posint;

    if nops(_self:-list_of_possible_cases) = 0 then
        return FAIL;
    fi;

    #keep trying all possible cases, starting from case 1 to case 3.
    #until one case succeed or all are tried

    for current_case_number in _self:-list_of_possible_cases do   
        sol := _self:-dsolve_case(current_case_number);
        if sol<>FAIL then 
            _self:-case_used_to_solve:=current_case_number;
            return sol; 
        fi;
    od;

    return FAIL;

end proc;

#--------------------------------------------------------------------
#  Called to solve the ode using a specific case number
#  made public to allow user to solve using specific case
#--------------------------------------------------------------------
export dsolve_case::static:=proc(_self,case_number::posint,$)
    local sol;

    if case_number>3 then ERROR("Only case number 1,2,3 are allowed"); fi;

    if nops(_self:-list_of_possible_cases)=0 then
        ERROR("No possible cases detected for this ode");
    fi;

    if not member(case_number,_self:-list_of_possible_cases) then
        ERROR("Case ", case_number, 
           " not one of possible cases ", _self:-list_of_possible_cases);
    fi;                

    if case_number = 1 then
       sol:= _self:-solve_case_1();   
    elif case_number = 2 then
       sol:= _self:-solve_case_2();
    else
       sol:= _self:-solve_case_3();
    fi;

    _self:-case_used_to_solve := case_number;
    return sol;
    
end proc;
#--------------------------------------------------------------------
# returns back the z''(x) = r(x) z(x) ode, which is the one
# actually solved by Kovacic algorithm
#--------------------------------------------------------------------
export get_z_ode::static:=proc(_self,$)
    local x := _self:-x;
    local r := _self:-r;
    local z := _self:-z;

    if _self:-s = 0 then
         return diff(z(x),x$2) = 0;
    else
         return diff(z(x),x$2) = numer(r)%/denom(r) * z(x);
    fi;
end proc;

#--------------------------------------------------------------------
# returns back the r term in the z''(x) = r(x) z(x) ode
#--------------------------------------------------------------------
export get_r::static:=proc(_self,$)
    if _self:-s = 0 then
        return 0;
    else
        return numer(_self:-r)%/denom(_self:-r);
    fi;
end proc;

#--------------------------------------------------------------------
#  returns s, where r = s/t and z'' = r z(t)
#--------------------------------------------------------------------
export get_s::static:=proc(_self,$)
    return _self:-s;
end proc;

#--------------------------------------------------------------------
#  returns t, where r = s/t and z'' = r z(t)
#--------------------------------------------------------------------
export get_t::static:=proc(_self,$)
    return _self:-t;
end proc;

#--------------------------------------------------------------------
# Returns back the original ode ( A y''+ B y' + C y = 0 )
#--------------------------------------------------------------------
export get_y_ode::static:=proc(_self,$)
    return _self:-original_ode;
end proc;

#--------------------------------------------------------------------
#  returns list of poles of r, where  z''(x) = r z(x)
#  The list has format [ [pole location,order], ...]
#--------------------------------------------------------------------
export get_poles::static:=proc(_self,$)
    return _self:-poles_list;
end proc;

#--------------------------------------------------------------------
#  returns O_infinity of r, where  z''(x) = r z(x)
#--------------------------------------------------------------------
export get_order_at_infinity::static:=proc(_self,$)
    return _self:-O_inf;
end proc;

#--------------------------------------------------------------------
#  returns list of possible kovacic cases possible.
#--------------------------------------------------------------------
export get_possible_cases::static:=proc(_self,$)::list;
    return _self:-list_of_possible_cases;
end proc;   

#--------------------------------------------------------------------
#  returns actual case number used when solving ode.
#  can be 1,2 or 3
#  if no cases applicable, then -1 is returned
#--------------------------------------------------------------------
export get_case_used::static:=proc(_self,$)::integer;    
    return _self:-case_used_to_solve;
end proc;   

#--------------------------------------------------------------------
#  returns n used for case 3. Either 4,6, or 12. 
#  if not case 3 used, then -1 is returned.
#--------------------------------------------------------------------
export get_n_case_3::static:=proc(_self,$)::integer;
    return _self:-n_used_for_case_3;
end proc;   


#--------------------------------------------------------------------
# All functions below are private
#--------------------------------------------------------------------

#--------------------------------------------------------------------
#This proc find all possible kovacic cases. These can be 1,2 or 3. 
#if none of these found, then empty list is returned, which is 
#case 4 in the paper. For example, if case 1 is only possible, 
#then [1] is returned. if case 1 and 2 are possible, then [1,2] 
#is returned, if all three cases possible, then [1,2,3] is returned.
#If no cases possible then [] returned.
#--------------------------------------------------------------------
local find_possible_cases::static:=proc(_self,$)
    local L::list := [];
    local poles_order := convert(_self:-poles_list[..,2],set);
    local T::set;

    #check for case 1
    T := select(Z-> Z>2,poles_order);
    if  nops( select(Z->type(Z,odd),T) )=0  and
        (_self:-O_inf<0 and type(_self:-O_inf,even)) or 
            _self:-O_inf=0 or _self:-O_inf>1 then
        L:= [1];
    fi;

    if nops(poles_order)>0 then #must have at least one pole for 2,3 cases

        if nops(poles_order)<>0 then #can not have pole order 0, i.e. no poles

            T:=select(Z-> Z>2,poles_order);

            # r must have at least one pole that is either odd order greater 
            #than 2 or else has order 2

            if  nops( select(Z->type(Z,odd),T) )<>0 
                     or member(2,poles_order) then
                L:= [ op(L),2 ];
            fi;

            #check for case 3

            if not member(0,poles_order) 
                   and nops(select(Z-> Z>2,poles_order))=0
                   and _self:-O_inf>=2 then

               L:= [ op(L),3 ];

            fi;
        fi;
    fi;

    _self:-list_of_possible_cases := L;

    return NULL;
end proc;

#--------------------------------------------------------------------
#called to find all poles of r and the order 
#of each pole. Uses Maple's sqrfree.
#--------------------------------------------------------------------
local generate_poles_and_order_list::static:=proc(_self,$)
    local L::list := [];
    local sol::list;
    local current_sol;
    local poles_list::list;
    local current_pole;
    local r := _self:-r, x := _self:-x;

    poles_list := sqrfree(denom(r),x); 
    poles_list := poles_list[2,..]; #we do not need the overall factor 

    if nops(poles_list) = 0 then
        _self:-poles_list:=[];
    else
        for current_pole in poles_list do 

            sol := solve(current_pole[1]=0,[x]);
            sol := ListTools:-Flatten(sol);

            for current_sol in sol do
                L := [op(L), [rhs(current_sol) ,current_pole[2] ] ]; 
            od;

        od; 

        _self:-poles_list := L;            
    fi;

    return NULL;
     
end proc:
#--------------------------------------------------------------------
#
#       C A S E     O N E      I M P L E M E N T A T I O N
#  
#  returns ode solution using case 1, or FAIL is no solution exist
#--------------------------------------------------------------------
local solve_case_1::static:=proc(_self,$)
    local O_infinity_set::kovacic_class:-case_one_O_inf;
    local gamma_set::set(kovacic_class:-case_one_gamma_entry):={}; 

    gamma_set, O_infinity_set := _self:-case_1_step_1();
    return _self:-case_1_step_2(gamma_set, O_infinity_set);
end proc;

#--------------------------------------------------------------------
# called from _self:-solve_case_1()
#--------------------------------------------------------------------
local case_1_step_1::static:=proc(_self,$)::
          set(kovacic_class:-case_one_gamma_entry),
          kovacic_class:-case_one_O_inf;

    local current_pole;
    local e::kovacic_class:-case_one_gamma_entry;
    local o::kovacic_class:-case_one_O_inf;
    local b,a,v,i::integer;
    local b_coeff_in_r,b_coeff_in_r_inf_square;
    local x := _self:-x, r := _self:-r;
    local N::integer;  
    local laurent_c;   
    local current_term;
    local b_from_r,b_from_laurent_series;

    #this contains all information generated for each pole of r
    #this is what is called the set GAMMA in the paper and 
    #in the diagram of algorithm above.
    local gamma_set::set(kovacic_class:-case_one_gamma_entry) := {}; 

    if nops(_self:-poles_list) = 0 then
        gamma_set := {};
    else
        for current_pole in _self:-poles_list do
            e := Object(kovacic_class:-case_one_gamma_entry);

            e:-pole_location  := current_pole[1];
            e:-pole_order     := current_pole[2];

            if e:-pole_order =1 then

               e:-sqrt_r      := 0;
               e:-alpha_plus  := 1;
               e:-alpha_minus := 1;
               gamma_set      := { op(gamma_set), e };

            elif e:-pole_order = 2 then

                e:-sqrt_r := 0;
                e:-b      := _self:-b_partial_fraction(r,x,e:-pole_location,2);
                e:-alpha_plus  := 1/2+1/2*sqrt(1+4*e:-b);
                e:-alpha_minus := 1/2-1/2*sqrt(1+4*e:-b);;
                gamma_set := { op(gamma_set), e };

            else

                v         := e:-pole_order/2;
                e:-sqrt_r := 0;

                for N from 2 to v  do
                    laurent_c := _self:-laurent_coeff(sqrt(r),x,
                                                      e:-pole_location,v,N);
                    current_term := laurent_c/(x-e:-pole_location)^N;
                    e:-sqrt_r    := e:-sqrt_r + current_term; 
                    if N = v then
                        a := laurent_c;
                    fi;
                od;

                b_from_r :=_self:-b_partial_fraction(r,x,e:-pole_location,v+1);
                b_from_laurent_series :=_self:-laurent_coeff(sqrt(r),x,
                                                    e:-pole_location,v,v+1);

                e:-b := b_from_r - b_from_laurent_series;
     
                e:-alpha_plus  := 1/2*((e:-b)/a  + v);
                e:-alpha_minus := 1/2*(-(e:-b)/a + v);
                gamma_set      := { op(gamma_set), e };

            fi;
        od; 
    fi;

    o := Object(case_one_O_inf);

    if _self:-O_inf > 2 then

        o:-sqrt_r_inf      := 0;
        o:-alpha_plus_inf  := 0;
        o:-alpha_minus_inf := 1;

    elif _self:-O_inf=2 then

        o:-sqrt_r_inf :=0;
        o:-b          := lcoeff(_self:-s) / lcoeff(_self:-t);
        b             := radsimp((1+4*o:-b)^(1/2));
        o:-alpha_plus_inf  := 1/2 + 1/2*b;
        o:-alpha_minus_inf := 1/2 - 1/2*b;

    else #order at infinity -2*v<= 0 which must be even

        v     := (-_self:-O_inf) / 2;  
        o:-sqrt_r_inf :=0;

        for i from 0 to v do

            laurent_c     := _self:-laurent_coeff(sqrt(r),x,infinity,v,i);    
            o:-sqrt_r_inf := o:-sqrt_r_inf + laurent_c*x^i;
            if i = v then
               o:-a := laurent_c;
            fi;

        od;
             
        b_coeff_in_r_inf_square := _self:-laurent_coeff(
                                            o:-sqrt_r_inf^2,x,0,v,v-1);    

        b_coeff_in_r       := _self:-get_coefficient_of_r(v-1);
        o:-b               := b_coeff_in_r - b_coeff_in_r_inf_square;
        o:-alpha_plus_inf  := 1/2*(  (o:-b)/(o:-a) - v); 
        o:-alpha_minus_inf := 1/2*( -(o:-b)/(o:-a) - v); 
    fi;

    return gamma_set,o;    

end proc;

#--------------------------------------------------------------------
# Finds b coefficient in r for the case one only when v<=0 
# using long division
#--------------------------------------------------------------------
local get_coefficient_of_r::static:=proc(_self,the_degree,$)
    local r := _self:-r;
    local x := _self:-x;
    local c,R,t;

    try
        c := coeff(r,x,the_degree);
    catch:
        R := rem(numer(r),denom(r),x);
        t := denom(r);
        c := lcoeff(R)/lcoeff(t);
    end try;

    return c;
end proc;

#--------------------------------------------------------------------
# called from _self:-solve_case_1()
# This determines the set of d non-negative integers and
# corresponding w for each d.
#--------------------------------------------------------------------
local case_1_step_2::static:=proc(_self,
           gamma_set::set(kovacic_class:-case_one_gamma_entry),
           O_infinity::kovacic_class:-case_one_O_inf,$)

    local d,N,K,number_of_poles::integer,current_alpha_infinity;
    local item::list;
    local the_sign::integer,sign_list::list;
    local w;
    local r_solution,y_solution;
    local x := _self:-x;

    #this will contains all data found for any nonnegative d. Each
    #entry will be a list of this form  
    #       [d, w]

    local good_d_found := Array(1..0);
    local B::Matrix; #this will contain good_d_found but as matrix

    number_of_poles := nops(gamma_set);
    sign_list := combinat:-permute([1$number_of_poles,-1$number_of_poles],
                                   number_of_poles);

    for K,current_alpha_infinity in 
                [O_infinity:-alpha_plus_inf,O_infinity:-alpha_minus_inf] do

        for item in sign_list do  

            d := 0;

            if number_of_poles>0 then

                for N,the_sign in item do
                    if the_sign = -1 then 
                        d := d-gamma_set[N]:-alpha_minus;
                    else
                        d := d-gamma_set[N]:-alpha_plus;
                    fi;                   
                od;

            fi;

            d := simplify(d + current_alpha_infinity);

            if type(d,'integer') and d >= 0 then
                w := 0;

                for N,the_sign in item do

                    if number_of_poles>0 then

                        if the_sign = -1 then 
                            w := w +(-1)*gamma_set[N]:-sqrt_r + 
                                    (gamma_set[N]:-alpha_minus)/
                                          (x - gamma_set[N]:-pole_location);
                        else
                            w := w + gamma_set[N]:-sqrt_r  + 
                                    (  gamma_set[N]:-alpha_plus)/
                                         (x  - gamma_set[N]:-pole_location);
                        fi;                   

                    fi;

                od;   

                #this to get the sign in according to paper. First + then -
                if K = 1 then
                   w := w + O_infinity:-sqrt_r_inf;
                else
                   w := w - O_infinity:-sqrt_r_inf;
                fi;

                good_d_found ,= [ d, w];
            fi;
        od;
    od;

    if numelems(good_d_found) = 0 then return FAIL; fi;

    #now convert the array to Matrix, and sort on d, so we 
    #start will the smallest
    #degree d, which is the first column, as that will be most efficient.
    #convert to set first, to remove any possible duplicat entries
    #then convert to Matrix

    convert(good_d_found,set);
    B := convert(convert(%,list),Matrix);
    B := B[sort(B[.., 1], 'output'= 'permutation')];

    for N from 1 to LinearAlgebra:-RowDimension(B) do

        r_solution := _self:-case_1_step_3(B[N,1], B[N,2]); #(d,w)  

        if r_solution <> FAIL then  
            y_solution := _self:-build_y_solution_from_r_solution(r_solution); 
            return y_solution;                                                             
        fi;

    od;

    return FAIL;

end proc;

#--------------------------------------------------------------------
# called from _self:-case_1_step_2()
#--------------------------------------------------------------------
local case_1_step_3::static:=proc(_self,d::nonnegative,w,$)
    local x := _self:-x;
    local r := _self:-r;
    local p;
    local i::integer;
    local a::nothing;
    local eq;
    local coeff_sol;
    local tmp,W;
    local final_result;

    p := x^d;

    for i from d-1 by -1 to 0 do        
        p := p + a[i] * x^i;
    od;    

    #using original kovacis method. Not Smith.
    eq := simplify( diff(p,x$2)+2*w*diff(p,x)+(diff(w,x)+w^2-r)*p) = 0;

    if d = 0 then
        if not evalb(eq) then return FAIL; fi;
    else    
        #solve for coefficients        
        try 
           coeff_sol := timelimit(30,solve(
                                   identity(eq,x), [seq(a[i],i=0..d-1)])); 
        catch:
            return FAIL;
        end try;

        if nops(coeff_sol) = 0 then return FAIL; fi;

        tmp := map(evalb,coeff_sol[1]);
        if has(tmp,true) then  return FAIL; fi;

        p := eval(p,coeff_sol[1]); #to force a[i] solutions to update        
    fi;

    W := diff(p,x)/p + w;
    W := radsimp(W);  

    tmp := diff(W,x)+W^2;

    if evalb( tmp = r) or is(tmp = r) or simplify(tmp-r)=0 then #can be used
        try
            tmp := int(w, x);    
        catch:
            return FAIL;
        end try;

        if has(tmp,int) then return FAIL; fi;

        final_result := simplify(p*exp(tmp)); 
        if has(final_result,signum) or has(final_result,csgn) then
            final_result := p*exp(tmp); 
        fi;

        return final_result;              
    else
        return FAIL;
    fi;  

end proc;

#--------------------------------------------------------------------
#
#  C A S E    T W O   I M P L E M E N T A T I O N
#
#  returns ode solution using case 2, or FAIL is no solution exist
#--------------------------------------------------------------------
local solve_case_2::static:=proc(_self,$)

    local E_inf::set;
    local gamma_set::set(kovacic_class:-case_2_and_3_gamma_entry):={}; 

    gamma_set, E_inf := _self:-case_2_step_1();
    return _self:-case_2_step_2(gamma_set, E_inf );

end proc;

#--------------------------------------------------------------------
#
#--------------------------------------------------------------------
local case_2_step_1::static:=proc(_self,$)::
        set(kovacic_class:-case_2_and_3_gamma_entry),set;

    local current_pole;
    local e::kovacic_class:-case_2_and_3_gamma_entry;
    local E_inf::set;
    local b;
    local x := _self:-x;
    local r := _self:-r;

    #this contains all information generated for each pole of r
    #this is what is called the set GAMMA in the paper and in the diagram of
    #algorithm above.
    local gamma_set::set(kovacic_class:-case_2_and_3_gamma_entry) := {}; 

    for current_pole in _self:-poles_list do

        e := Object(kovacic_class:-case_2_and_3_gamma_entry);
        e:-pole_location  := current_pole[1];
        e:-pole_order     := current_pole[2];

        if e:-pole_order = 1 then

           e:-Ec     := {4};
           gamma_set := { op(gamma_set), e };

        elif e:-pole_order = 2 then

            e:-b      := _self:-b_partial_fraction(r,x,e:-pole_location,2);
            e:-Ec     := {2,2+2*sqrt(1+4* e:-b),2-2*sqrt(1+4* e:-b)};             
            e:-Ec     := select(z->type(z,integer),e:-Ec);
            gamma_set := { op(gamma_set), e };

        else

           e:-Ec     := {e:-pole_order}; 
           gamma_set := { op(gamma_set), e };

        fi;
    od; 

    if _self:-O_inf>2 then

        E_inf := {0,2,4};

    elif _self:-O_inf=2 then

        b     := lcoeff(_self:-s) / lcoeff(_self:-t);
        E_inf := {2,2+2*sqrt(1+4*b),2-2*sqrt(1+4*b)};  
        E_inf := select(z->type(z,integer),E_inf);

    else #order at infinity v< 2

        E_inf := {_self:-O_inf};  

    fi;

    return gamma_set,E_inf;    
end proc;

#--------------------------------------------------------------------
# called from _self:-solve_case_2()
# This determines the set of d non-negative integers and
# corresponding w for each d.
#--------------------------------------------------------------------
local case_2_step_2::static:=proc(_self,
           gamma_set::set(kovacic_class:-case_2_and_3_gamma_entry),
           E_inf::set,
           $)

    local L::list := [];
    local item,current_E_inf;
    local d,N,ee,theta;
    local x := _self:-x;
    local r_solution,y_solution;

    #this will contains all data found for any nonnegative d. Each
    #entry will be a list of this form  
    #       [d, theta]
    #so if we obtain say 3 values of d that are nonnegative, 
    #there will be 3 such lists in this array

    local good_d_found := Array(1..0);
    local B::Matrix; #this will contain good_d_found as matrix

    for item in gamma_set do
        L := [ op(L), convert(item:-Ec,list) ];
    od;

    #now find all possible tuples
    if nops(L)>1 then
       L := kovacic_class:-cartProdSeq(op(L));
    else
       L := L[1];
    fi;

    for current_E_inf in E_inf do
        for item in L do            

            d := 1/2*( current_E_inf - add(item));

            if type(d,'integer') and d >= 0 then      
                theta :=0;

                for N,ee in item do
                    theta := theta + ee/(x-gamma_set[N]:-pole_location);                
                od;

                theta         := 1/2*theta;
                good_d_found ,= [ d, theta];                
            fi;

        od;
    od;

    if numelems(good_d_found) = 0 then return FAIL; fi;

    #now convert the array to Matrix, and sort on d, so 
    #we start will the smallest degree d, which is the first column, as 
    #that will be most efficient.
    #convert to set first, to remove any possible duplicat entries
    #then convert to Matrix

    convert(good_d_found,set);
    B := convert(convert(%,list),Matrix);
    B := B[sort(B[.., 1], 'output'= 'permutation')];

    for N from 1 to LinearAlgebra:-RowDimension(B) do

        r_solution := _self:-case_2_step_3(B[N,1], B[N,2]);  #(d,theta)                                                 

        if r_solution <> FAIL then  
            y_solution := _self:-build_y_solution_from_r_solution(r_solution); 
            return y_solution;                                                             
        fi;

    od;

    return FAIL;

end proc;

#--------------------------------------------------------------------
# called from _self:-case_2_step_2()
#--------------------------------------------------------------------
local case_2_step_3::static:=proc(_self,d::nonnegative,theta,$)
 
    local p, i;
    local a::nothing;
    local tmp;
    local coeff_sol := [];
    local eq;
    local phi;
    local sol_w;
    local current_w;
    local r:=_self:-r;
    local w::nothing;
    local x := _self:-x;
    local sol;

    p := x^d;
        
    for i from d-1 by -1 to 0 do        
        p := p + a[i] * x^i;
    od;    

    eq:= simplify(diff(p,x$3) + 3*theta*diff(p,x$2) + 
            (3*theta^2+3*diff(theta,x) -4*r) * diff(p,x) 
              + ( diff(theta,x$2)+3*theta*diff(theta,x)+theta^3 - 
                 4*r*theta - 2*diff(r,x))*p) = 0;
           
    if d=0 then
        if not evalb(eq) then return FAIL; fi;
    else                   
        #solve for polynomial coefficients
        try
            coeff_sol:= timelimit(30,solve( 
                                   identity(eq,x), [seq(a[i],i=0..d-1)]));
        catch:
            return FAIL;
        end try;

        if nops(coeff_sol) = 0 then return FAIL; fi;
        
        tmp := map(evalb,coeff_sol[1]);
        if has(tmp,true) then return FAIL; fi;

        p := eval(p,coeff_sol[1]); #to force a[i] solutions to update        
    fi;        

    phi := theta + diff(p,x)/p;        
    eq  := w^2 - phi*w +  simplify(1/2*diff(phi,x)+1/2*phi^2-r) = 0;   

    try
        sol_w := timelimit(30,solve(eq,[w])); #changed to []
        sol_w := ListTools:-Flatten(sol_w);
    catch:
        return FAIL;
    end try;

    if nops(sol_w) = 0 then return FAIL; fi;

    for current_w in sol_w do

        current_w := radsimp(rhs(current_w));                

        #verify w before using it. Added 1/12/2022 4 PM
        tmp := diff(current_w,x)+current_w^2;
        if evalb( tmp= r) or is(tmp = r) or simplify(tmp-r)=0 then
            try
               sol := timelimit(30,int(current_w, x));  
            catch:
               return FAIL;
            end try;

            if has(sol,int) then return FAIL; fi;

            return simplify(exp(sol)); 
        fi;         

    od;

end proc;

#--------------------------------------------------------------------
#
#  C A S E   T H R E E    I M P L E M E N T A T I O N
#
#  returns ode solution using case 3, or FAIL is no solution exist
#--------------------------------------------------------------------

local solve_case_3::static:=proc(_self,$)

    local E_inf::set;
    local gamma_set::set(kovacic_class:-case_2_and_3_gamma_entry):={}; 
    local sol;

    #these are possible degress of w to try until one works or none works
    #local w_degree::list := [4,6,12]; 
    local w_degree::list := [4,6,12]; 
    local n::posint;

    for n in w_degree do
        gamma_set, E_inf := _self:-case_3_step_1(n);
        sol              := _self:-case_3_step_2(gamma_set, E_inf, n );

        if sol<>FAIL then return sol; fi;
    od;

    return FAIL;

end proc;

#--------------------------------------------------------------------
# First step in case 3
#--------------------------------------------------------------------
local case_3_step_1::static:=proc(_self,
           n::posint,
           $)::set(kovacic_class:-case_2_and_3_gamma_entry),set;


    local current_pole;
    local e::kovacic_class:-case_2_and_3_gamma_entry;
    local E_inf::set;
    local b,k;
    local x := _self:-x;
    local r := _self:-r;

    #this contains all information generated for each pole of r
    #this is what is called the set GAMMA in the paper and in the diagram of
    #algorithm above.
    local gamma_set::set(kovacic_class:-case_2_and_3_gamma_entry):={}; 

    for current_pole in _self:-poles_list do

        e := Object(kovacic_class:-case_2_and_3_gamma_entry);

        e:-pole_location := current_pole[1];
        e:-pole_order    := current_pole[2];

        if e:-pole_order = 1 then           

           e:-Ec     := {12};
           gamma_set := { op(gamma_set), e };

        elif e:-pole_order = 2 then

            e:-b      := _self:-b_partial_fraction(r,x,e:-pole_location,2);
            e:-Ec     := {seq( 6+12*k/n*sqrt(1+4*(e:-b)),k=-n/2..n/2,1)};        
            e:-Ec     := select(z->type(z,integer),e:-Ec);
            gamma_set := { op(gamma_set), e };

        else
           ERROR("Internal error. Case 3 can only have poles of order 1 or 2");
        fi;
    od; 

    #same formula, but different b
    b     := lcoeff(_self:-s) / lcoeff(_self:-t);    
    E_inf := {seq( 6+12*k/n*sqrt(1+4*b),k=-n/2..n/2,1)};    
    E_inf := select(z->type(z,integer),E_inf);

    return gamma_set,E_inf;    

end proc;

#--------------------------------------------------------------------
# Second step in case 3
#
# called from _self:-solve_case_3()
# This determines the set of d non-negative integers and
# corresponding w for each d.
#--------------------------------------------------------------------
local case_3_step_2::static:=proc(
        _self,
        gamma_set::set(kovacic_class:-case_2_and_3_gamma_entry),
        E_inf::set,
        n::posint,$)

    local L::list := [];
    local item,current_E_inf;
    local d,ee,theta,N;
    local x := _self:-x;
    local r_solution,y_solution;
    local S;
    local current_iteration::integer;
    local tmp;

    #this will contains all data found for any nonnegative d. Each
    #entry will be a list of this form  
    #       [d, theta, S]
    #so if we obtain say 3 values of d that are nonnegative
    local good_d_found := Array(1..0);
    local B::Matrix; #this will contain good_d_found as matrix 

    #DEBUG();
    for item in gamma_set do
        L := [ op(L), convert(item:-Ec,list) ];
    od;

    #now find all possible tuples
    if nops(L)>1 then
       L := kovacic_class:-cartProdSeq(op(L));
    else
       L := L[1];
    fi;

    current_iteration:=0;
    for current_E_inf in E_inf do
        for item in L do            
            current_iteration := current_iteration+1;

            d := n/12*( current_E_inf - add(item));
           
            if type(d,'integer') and d >= 0 then
                theta :=0;                
                for N,ee in item do
                    theta := theta + ee/(x-gamma_set[N]:-pole_location);                                    
                od;
                theta := n/12*theta;
                theta := simplify(theta);
                S := mul( (x-gamma_set[N]:-pole_location), N=1..nops(item));
                tmp := simplify(S) assuming real;
                if not has(tmp,csgn) and not has(tmp,signum) then
                   S:= tmp;
                fi;

                good_d_found ,= [ d, theta, S];
            fi;
        od;
    od;

    #now convert the array to Matrix, and sort on d, so we start
    # with the smallest
    #degree d, which is the first column, as that will be most efficient.

    if numelems(good_d_found) = 0 then return FAIL; fi;

    #convert to set first, to remove any possible duplicat entries
    #then convert to Matrix
    convert(good_d_found,set);
    B := convert(convert(%,list),Matrix);
    B := B[sort(B[.., 1], 'output'= 'permutation')];

    for N from 1 to LinearAlgebra:-RowDimension(B) do
        #(d,theta,S,n)
        r_solution := _self:-case_3_step_3(B[N,1], B[N,2], B[N,3],n); 

        if r_solution <> FAIL then  
            _self:-n_used_for_case_3 := n;
            y_solution := _self:-build_y_solution_from_r_solution(r_solution); 
            return y_solution;                                                             
        fi;
    od;

    return FAIL;
end proc;

#--------------------------------------------------------------------
# Third and final step in case 3.
# called from _self:-case_3_step_2(). Returns solution or FAIL is no
# solution found.
#--------------------------------------------------------------------
local case_3_step_3::static:=proc(
                 _self,
                 d::nonnegative,
                 theta,
                 S,
                 n::posint,
                 $)
 
    local p, i,P_minus_1;
    local a::nothing;
    local final_result,tmp;
    local coeff_sol := [];
    local sol_w;
    local current_w;
    local r := _self:-r;
    local x := _self:-x;
    local omega::symbol;
    local P := Array(-1..n);
    local result_of_simplify;
    local result_of_is_check;
    local omega_equation;

    #this makes p(x). For example if d=3, then the result will be 
    #         p(x) = x^3 + a(2) x^2 + a(1) x  + a(0) 
    #where the number of unknowns to determine is always the same
    #as the degree, in this case a(0),a(1),a(2)    

    p := x^d; #this will be 1 if degree is zero

    for i from d-1 by -1 to 0 do        
        p := p + a[i] * x^i;
    od;    

    #build the P_n polynomials based on p(x) above
    P[n] := -p;

    for i from n by -1 to 0 do

        if i=n then
            P[i-1] := -S * diff(P[i], x) - S*theta*P[i];
            P[i-1] := simplify( P[i-1]);
        else
            P[i-1] := -S * diff(P[i], x) + 
                  ( (n-i)*diff(S,x) - S*theta)*P[i] - (n-i)*(i+1)*S^2*r*P[i+1];
            P[i-1] := simplify( P[i-1]);
        fi;

    od;

    #solve P[-1] for p(x) if needed.
    P_minus_1 := expand(numer(radsimp(P[-1]))); 

    if P_minus_1 <>0 and evalb(p<>1) then
        if not hastype(P_minus_1,'indexed')  then #it must be indexed now
           ERROR("Internal error. Please report. This should not happen");
        fi;

        try
            coeff_sol := timelimit(30,solve(identity(P_minus_1,x), [seq(a[i],i=0..d-1)])); 
        catch:
            return FAIL;
        end try;
        
        if nops(coeff_sol)=0 then #unable to solve
           return FAIL;
        fi;
            
        map(evalb,coeff_sol[1]); #check all solved for
        if has(%,true) then return FAIL; fi;            
    fi;
       
    #build the equation for omega
    omega_equation := 0;

    for i from 0 to n do
        omega_equation := omega_equation + S^i*P[i]/(n-i)! * omega^i ;
    od;    
        
    omega_equation := simplify(omega_equation);    

    if nops(coeff_sol)<>0 then
        #to force a[i] solutions to update to solved coefficients 
        omega_equation := eval(omega_equation,coeff_sol[1]); 
    fi;

    try
        sol_w := timelimit(30, solve(omega_equation=0, [omega]));
        sol_w := ListTools:-Flatten(sol_w);
    catch:
        return FAIL;    
    end try;

    if nops(sol_w) = 0 then return FAIL; fi;    

    #go over each w solution and use one that works. verify before using

    for current_w in sol_w do

        current_w := rhs(current_w);

        if not has(current_w, RootOf) then               
            try 
                current_w := timelimit(30,simplify(current_w));                
            catch:
                NULL;
            end try;

            if is_w_verified(current_w,x,r) then
                final_result := _self:-simplify_final_result(current_w,x);            
                if final_result<>FAIL then
                   return final_result;              
                fi;                
            fi;
        else #no Rootof, try to resolve       
            try
                current_w := timelimit(30,[allvalues(current_w)]);
            catch:
                return FAIL;
            end try;
                           
            current_w := current_w[1]; #just use any root. Pick first

            if not has(current_w, RootOf) then
                try
                    current_w := timelimit(30,simplify(current_w));                
                catch:
                    NULL;
                end try;

                if is_w_verified(current_w,x,r) then
                    final_result := _self:-simplify_final_result(current_w,x);                                
                    if final_result<>FAIL then
                       return final_result;              
                    fi;

                fi;        
            fi;
        fi;
    od;

    return FAIL;

end proc;


#--------------------------------------------------------------------
# Called from case 3, step 3 to verify w
#--------------------------------------------------------------------
local is_w_verified:=proc(current_w,x,r)::truefalse;
    local tmp;
    local result_of_is_check::truefalse;
    local result_of_simplify;

    tmp := diff(current_w,x)+current_w^2;

    if evalb( tmp=r) then return true; fi;
    
    try
        result_of_simplify := timelimit(30,simplify(tmp-r));
        if evalb(result_of_simplify=0) then return true; fi;
    catch:
        NULL;
    end try;

    try
        result_of_is_check := timelimit(30,is(tmp = r));
        if result_of_is_check then return true; fi;
    catch:
        NULL;
    end try;                
    
    
    return false;
end proc;

#--------------------------------------------------------------------
# Called from case 3, step 3 to simplify final result
#--------------------------------------------------------------------

local simplify_final_result::static:=proc(_self,omega,x,$)
    local final_result;
    local integral_result;

    try
        integral_result := timelimit(30,int(omega, x));                            
    catch:            
        return FAIL;           
    end try;

    if has(integral_result,int) or has(integral_result,Int) then 
        return exp(Int(omega, x));
    fi;

    try
        final_result := timelimit(30,simplify(exp(integral_result))) assuming real; 
        if has(final_result,signum) or has(final_result,csgn) 
            or has(final_result,abs) then
            final_result := exp(integral_result);
        fi;
    catch:
        final_result:= exp(integral_result);
    end try;
                            
    return final_result;            
end proc;


#--------------------------------------------------------------------
# External proc helper
# provided thanks to Joseph Riel
#--------------------------------------------------------------------
local cartProdSeq:= proc(L::seq(list))
    local Seq::nothing,i::nothing,j;
    option `Copyright (C) 2007, Joseph Riel. All rights reserved.`;
        eval([subs(Seq= seq, foldl(Seq, [cat(i, 1..nargs)], 
                                  seq(cat(i,j)= L[j], j= nargs..1, -1)))])
end proc:

#--------------------------------------------------------------------
# Checks for special math function.
# This function was provided thanks to Carl Love.
# modified to allow some special functions.
#--------------------------------------------------------------------
local has_special_math_function:= subs(
    _F= {(op@FunctionAdvisor)~(FunctionAdvisor("class_members", "quiet"), "quiet")[]}
        minus (   {FunctionAdvisor("elementary", "quiet")[]}
                  union {erf,erfc, erfi, Im,Re,signum,max,argument} ),
    proc(e::algebraic, x::{name, set(name)}:= {})
        hastype(e, And(specfunc(_F), dependent(x)))
    end proc
):

#--------------------------------------------------------------------
# called if step 3 is successful. Build y solution from r solution
#--------------------------------------------------------------------
local build_y_solution_from_r_solution::static:=proc(_self,r_solution,$)

    local y1,y2;
    local int_B_over_A;
    local A :=_self:-A, B:=_self:-B;
    local x :=_self:-x, y:= _self:-y;
    local tmp;

    if B = 0 then
       y1 := r_solution;
       int_B_over_A := 0;
    else
        try
            int_B_over_A := timelimit(60,int(-B/A,x));        
            if has(int_B_over_A,int) or has(int_B_over_A,RootOf) then                            
                int_B_over_A := Int(-B/A, x);
                y1 :=r_solution*exp(1/2*int_B_over_A);
            else
                y1 := simplify(r_solution*exp(1/2*int_B_over_A)); 
                if has(y1,signum) or has(y1,csgn) then                            
                    y1 := r_solution*exp(1/2*int_B_over_A);
                fi;
            fi;
        catch:
            int_B_over_A := Int(-B/A,x);
            y1 := r_solution*exp(1/2*int_B_over_A);
        end try;                        
    fi;

    try 
        y2 := timelimit(30, int( simplify(exp(int_B_over_A))/y1^2,x));

        if kovacic_class:-has_special_math_function(y2,x) then
            y2 := y1 * Int( exp(int_B_over_A)/y1^2,x)    
        else
            y2 := y1 * y2;
        fi;
        tmp := simplify(y2);
        if has(tmp,signum) or has(tmp,csgn) then                            
            y2 := y1 * y2;
        else
            y2 := tmp;
        fi;
    catch:
        y2 := y1 * Int( exp(int_B_over_A)/y1^2,x)    
    end try;
        
    return y(x) = _C1*y1 + _C2*y2;

end proc;

#--------------------------------------------------------------------
# Helper private function called to obtain b from the partial fraction
# decomposition of r needed by the differenent case implementations
#--------------------------------------------------------------------
local b_partial_fraction::static:=proc(_self,
        r,
        x::symbol,
        pole_location,
        the_power::posint,$)

    local r_partial_fraction,T::nothing;

    r_partial_fraction := allvalues(convert(r,'fullparfrac',x));

    #adding dummy T so that select below always works
    r_partial_fraction := T+r_partial_fraction; 
    select(z->hastype(z,
            anything/(anything*identical(x - pole_location)^the_power)),
            r_partial_fraction);

    return coeff(%,1/(x-pole_location)^the_power);

end proc;

#--------------------------------------------------------------------
# Helper private function called to return Laurent series coefficient
# of the 1/(x-c)^n  term. where n=1,2,3,.... for the function f(x)
# expandid about pole of order m
# if c=infinity, then x is replaced by 1/y and expansion is around 0
# for infinity, n=0,1,2,...
#--------------------------------------------------------------------
local laurent_coeff::static:=proc(_self,
        f,
        x::symbol,
        c,
        m::integer, 
        n::integer,$)

    local the_coeff,y::nothing,fy;

    if c = infinity then
        fy := eval(f,x=1/y);
        if m-n>0 then
           the_coeff := limit( diff( y^m * fy,y$(m-n)),y=0,right)/(m-n)!;
        elif m-n=0 then
            the_coeff := limit( y^m * fy,y=0,right)/(m-n)!;
        else
           the_coeff := 0;
        fi;
    else    
        if m-n>0 then
            the_coeff := limit( diff( (x-c)^m * f,x$(m-n)),x=c,right)/(m-n)! ;
        elif m-n=0 then
           the_coeff := limit( (x-c)^m * f,x=c,right)/(m-n)!;
        else
           the_coeff := 0;
        fi;
    fi;

    the_coeff := eval(the_coeff,[csgn=1,signum=1]);

    return the_coeff;

end proc;

end module;
\end{lstlisting}


\begin{thebibliography}{1}
 
\bibitem{Kovacic86}
Jerald J. Kovacic.
\newblock An Algorithm for Solving Second Order Linear Homogeneous Differential Equations.
\newblock {\em J. Symb. Comput.}, 2(1):3--43, 1986.
%
\bibitem{Smith84}
Carolyn J. Smith.
\newblock A DISCUSSION AND IMPLEMENTATION OF KOVACICS ALGORITHM FOR ORDINARY DIFFERENTIAL EQUATIONS.
\newblock {\em Research Report CS-84-35}, October 1984.
%
\bibitem{Saunders}
B. David Saunders.
\newblock An Implementation of Kovacic's Algorithm for Solving Second Order Linear Homogeneous Differential Equations.
\newblock {\em Proceedings of the 1981 ACM Symposium on Symbolic and Algebraic Computation}, 1981.
%
\end{thebibliography}
\end{document}